\PassOptionsToPackage{unicode}{hyperref}
\PassOptionsToPackage{hyphens}{url}
\PassOptionsToPackage{dvipsnames,svgnames,x11names}{xcolor}
\documentclass[
  a4paper,
]{article}

\usepackage{amsmath,amssymb}
\usepackage{lmodern}
\usepackage{setspace}
\usepackage{iftex}
\ifPDFTeX
  \usepackage[T1]{fontenc}
  \usepackage[utf8]{inputenc}
  \usepackage{textcomp} 
\else 
  \usepackage{unicode-math}
  \defaultfontfeatures{Scale=MatchLowercase}
  \defaultfontfeatures[\rmfamily]{Ligatures=TeX,Scale=1}
\fi
\IfFileExists{upquote.sty}{\usepackage{upquote}}{}
\IfFileExists{microtype.sty}{
  \usepackage[]{microtype}
  \UseMicrotypeSet[protrusion]{basicmath} 
}{}
\makeatletter
\@ifundefined{KOMAClassName}{
  \IfFileExists{parskip.sty}{%
    \usepackage{parskip}
  }{
    \setlength{\parindent}{0pt}
    \setlength{\parskip}{6pt plus 2pt minus 1pt}}
}{
  \KOMAoptions{parskip=half}}
\makeatother
\usepackage{xcolor}
\ifx\paragraph\undefined\else
  \let\oldparagraph\paragraph
  \renewcommand{\paragraph}[1]{\oldparagraph{#1}\mbox{}}
\fi
\ifx\subparagraph\undefined\else
  \let\oldsubparagraph\subparagraph
  \renewcommand{\subparagraph}[1]{\oldsubparagraph{#1}\mbox{}}
\fi

\usepackage{longtable,booktabs,array}
\usepackage{calc} 
\usepackage{etoolbox}
\makeatletter
\patchcmd\longtable{\par}{\if@noskipsec\mbox{}\fi\par}{}{}
\makeatother
\IfFileExists{footnotehyper.sty}{\usepackage{footnotehyper}}{\usepackage{footnote}}
\makesavenoteenv{longtable}
\usepackage{graphicx}
\makeatletter
\def\maxwidth{\ifdim\Gin@nat@width>\linewidth\linewidth\else\Gin@nat@width\fi}
\def\maxheight{\ifdim\Gin@nat@height>\textheight\textheight\else\Gin@nat@height\fi}
\makeatother
\setkeys{Gin}{width=\maxwidth,height=\maxheight,keepaspectratio}
\makeatletter
\def\fps@figure{htbp}
\makeatother

\usepackage[utf8]{inputenc}
\usepackage{verbatim}
\usepackage{amsfonts}
\usepackage[nonumberlist]{glossaries}
\usepackage{amssymb,amsmath,amsthm}
\usepackage{array, booktabs, makecell}
\usepackage{graphicx}
\usepackage{appendix}
\usepackage{authblk}
\usepackage{setspace}
\usepackage{caption}
\usepackage{dsfont}
\usepackage{xcolor}
\usepackage{lipsum}
\usepackage{multicol}
\usepackage{multirow}
\usepackage{subcaption}
\usepackage{natbib}
\usepackage{textcomp}
\usepackage{makecell}
\usepackage{adjustbox}
\usepackage{eurosym}
\usepackage{enumitem}
\usepackage{bm}
\usepackage{chngcntr}
\usepackage{colortbl}
\usepackage{pifont}
\usepackage{subfiles}
\usepackage{hyperref}
\usepackage{longtable}
\newcommand{\cmark}{\ding{51}}

\usepackage{booktabs}
\usepackage{longtable}
\usepackage{array}
\usepackage{multirow}
\usepackage{wrapfig}
\usepackage{float}
\usepackage{colortbl}
\usepackage{pdflscape}
\usepackage{tabu}
\usepackage{threeparttable}
\usepackage{threeparttablex}
\usepackage[normalem]{ulem}
\usepackage{makecell}
\usepackage{xcolor}
\usepackage{arxiv}
\usepackage{orcidlink}
\usepackage{amsmath}
\usepackage[T1]{fontenc}
\makeatletter
\@ifpackageloaded{caption}{}{\usepackage{caption}}
\AtBeginDocument{%
\ifdefined\contentsname
  \renewcommand*\contentsname{Table of contents}
\else
  \newcommand\contentsname{Table of contents}
\fi
\ifdefined\listfigurename
  \renewcommand*\listfigurename{List of Figures}
\else
  \newcommand\listfigurename{List of Figures}
\fi
\ifdefined\listtablename
  \renewcommand*\listtablename{List of Tables}
\else
  \newcommand\listtablename{List of Tables}
\fi
\ifdefined\figurename
  \renewcommand*\figurename{Figure}
\else
  \newcommand\figurename{Figure}
\fi
\ifdefined\tablename
  \renewcommand*\tablename{Table}
\else
  \newcommand\tablename{Table}
\fi
}
\@ifpackageloaded{float}{}{\usepackage{float}}
\floatstyle{ruled}
\@ifundefined{c@chapter}{\newfloat{codelisting}{h}{lop}}{\newfloat{codelisting}{h}{lop}[chapter]}
\floatname{codelisting}{Listing}

\makeatother
\makeatletter
\@ifpackageloaded{caption}{}{\usepackage{caption}}
\@ifpackageloaded{subcaption}{}{\usepackage{subcaption}}
\makeatother
\makeatletter
\@ifpackageloaded{tcolorbox}{}{\usepackage[many]{tcolorbox}}
\makeatother
\makeatletter
\@ifundefined{shadecolor}{\definecolor{shadecolor}{rgb}{.97, .97, .97}}
\makeatother
\ifLuaTeX
  \usepackage{selnolig}  
\fi
\usepackage[]{natbib}
\IfFileExists{bookmark.sty}{\usepackage{bookmark}}{\usepackage{hyperref}}
\IfFileExists{xurl.sty}{\usepackage{xurl}}{} 
\hypersetup{
  pdftitle={Analysing and visualising bike-sharing demand with outliers},
  pdfauthor={Nicola RENNIE; Catherine CLEOPHAS*; Adam M. SYKULSKI; Florian DOST},
  colorlinks=true,
  linkcolor={blue},
  filecolor={Maroon},
  citecolor={blue},
  urlcolor={Blue},
  pdfcreator={LaTeX via pandoc}}

\newcommand{\runninghead}{A Preprint }
\title{Analysing and visualising bike-sharing demand with outliers}
\author{
\textbf{Nicola RENNIE}\\STOR-i Centre for Doctoral Training\\Lancaster
University\\Lancaster,\ LA1
4YW\\\href{mailto:n.rennie@lancaster.ac.uk}{n.rennie@lancaster.ac.uk}\\\\\\\\
\textbf{Catherine CLEOPHAS*}\\Institute for
Business\\Christian-Albrechts-University
Kiel\\Kiel,\ 24118\\\href{mailto:cleophas@bwl.uni-kiel.de}{cleophas@bwl.uni-kiel.de}\\ * Corresponding author \\\\\\
\textbf{Adam M. SYKULSKI}\\Dept. of Mathematics\\Imperial College
London\\London,\ SW7
2AZ\\\href{mailto:adam.sykulski@imperial.ac.uk}{adam.sykulski@imperial.ac.uk}\\\\\\\\
\textbf{Florian DOST}\\Institute of Business and Economics\\Brandenburg
University of
Technology\\Cottbus,\ 03046\\\href{mailto:florian.dost@b-tu.de}{florian.dost@b-tu.de}\\}

\date{\vspace{-3ex}}
\begin{document}
\maketitle
\begin{abstract}
Bike-sharing is a popular component of sustainable urban mobility. It
requires anticipatory planning, e.g.~of station locations and inventory,
to balance expected demand and capacity. However, external factors such
as extreme weather or glitches in public transport, can cause demand to
deviate from baseline levels. Identifying such outliers keeps historic
data reliable and improves forecasts. In this paper we show how outliers
can be identified by clustering stations and applying a functional depth
analysis. We apply our analysis techniques to the Washington D.C.
Capital Bikeshare data set as the running example throughout the paper,
but our methodology is general by design. Furthermore, we offer an array
of meaningful visualisations to communicate findings and highlight
patterns in demand. Last but not least, we formulate managerial
recommendations on how to use both the demand forecast and the
identified outliers in the bike-sharing planning process.

\end{abstract}
\ifdefined\Shaded\renewenvironment{Shaded}{\begin{tcolorbox}[borderline west={3pt}{0pt}{shadecolor}, breakable, enhanced, interior hidden, boxrule=0pt, frame hidden, sharp corners]}{\end{tcolorbox}}\fi

\setstretch{1.1}
\hypertarget{sec-introduction}{%
\section{Introduction and Background}\label{sec-introduction}}

As a component of sustainable urban mobility, bike-sharing is on the
rise in cities around the world. Empirical research as exemplified by
\citet{teixeira2021empirical} and \citet{blazanin2022scooter} has been
dedicated to highlighting requirements and beneficial impacts of the
related modal shifts in urban transport. As pointed out in the related
references, careful planning is required to make so-called shared
micromobility systems more attractive than less environmentally friendly
alternatives, such as cars. While the concept of shared micromobility
also includes e-scooters and dockless bike-shares, here, we focus on
systems that rely on a set of dedicated stations. As, for example,
\citet{luo2019comparative} point out, while dockless systems offer more
convenience and equity to users, their CO2 footprint is significantly
higher due to the reduced lifespan of vehicles. On the other hand,
station-based sharing systems require carefully planned station
locations and well-balanced inventory levels to ensure adequate service
provision. When, for example, a station is mostly used to pick-up bikes,
re-balancing ensures a steady supply and avoids service denials.

Planning for bike-sharing operations means determining, for example, the
best distribution of stations across the service area
\citep{Ciancio2017}, the best distribution of bikes across stations
\citep{Zhu2021}, and the best path for truck drivers to take when
re-distributing bikes each day \citep{Schuijbroek2017}. When taking a
pro-active approach to planning, optimisation procedures that determine
stock levels per station rely on predicted demand. When taking a
re-active approach, quick online decision-making is crucial to maintain
a good service level. Since inefficient re-balancing operations are a
major cost driver for operators \citep{Schuijbroek2017}, identifying
demand outliers to improve efficiency in bike-sharing systems is highly
important. Unaccounted-for outliers can affect bike-sharing systems in
two ways: (i) outliers in historic data contaminate the forecasts used
in future inventory management, and (ii) on the day demand levels may
indicate that the schedule is non-optimal for the current day and
drivers should be re-routed.

Therefore, identification of outlier demand has several potential
benefits for bike-sharing forecasting and planning: 1) Detecting
outliers early in the day, through online analysis as proposed in
\citet{Rennie2021}, allows for rapid interventions to better re-allocate
bikes on a given day; 2) Removing any detected outliers from training
data for demand forecasting would improve results on predicting
reference demand curves; 3) If outliers can be attributed to specific
events previously unknown, extending future forecast models to include
such events can improve forecasts; 4) Even when explanatory factors for
outliers cannot be determined, if such outliers are concentrated
spatio-temporally in certain stations, this knowledge can better support
planning decisions; and 5) Identifying changes in the underlying
reference model when patterns in the detected outliers are observed can
trigger a review of the current forecasting method.

We define outlier demand as a short-term change in demand, resulting in
usage levels which deviate from \emph{regular} usage. Note that, to
count as an outlier, a demand shift has to exceed the general degree of
random variation observed in demand over time. In this paper, we focus
on demand observed at bike-sharing stations, as these are the target of
inventory rebalancing efforts. In contrast to other classical mobility
problems, such as those related to buses or trains, bike-sharing
capacity is on the vertices of the transport network, rather than on the
edges.

As an example of existing work in this area, \citet{NeumannSaavedra2021}
discuss the problem of variability in bike-sharing demand and propose a
rule-based method to adjust the redistribution plan when demand differs
from the forecast. In a simulation study, they show that service levels
can be improved when adjustments are made to the optimal redistribution
plan. Wider literature on outlier detection in transport planning is
scarce -- e.g., \citet{Rennie2021b} consider identifying and correcting
for outliers in revenue management systems in railways.

\citet{Talvitie1978} find that outliers can have a substantial effect on
the predictions of usage of different urban transport modes, but only
apply a simplistic trimming method to identify outliers. In the road
traffic domain, \citet{Guo2015} suggest a procedure for identifying
outliers in real time based on the conditional variance of predictions,
and determine that incorporating information on such outliers into
future predictions increases the systems performance.

Furthermore, as indicated, e.g., in \citet{basole2021visualization}, to
account for demand outliers and adjust planning, experts require
meaningful visualisations. Therefore, we propose a set of visualisations
to help identify and analyse spatial and temporal patterns in the
detected outliers. For example, given shifts in the availability of
urban infrastructure, a subset of stations may be predisposed to
outliers and as such, this area would be a good target for a temporary
``pop-up'' station. Throughout this paper, we assume that bike-sharing
companies employ analysts who are in charge of strategic decisions, such
as where to locate stations, tactical decisions, such as what number of
bikes should be available on any given day at those stations, and
operational decisions, such as on-the-fly rebalancing of bikes. In that,
we follow previous research, such as \citet{aswang2016modeling}, who
expect analysts to evaluate bike share programs and station locations,
or \citet{orma2021investigating}, who consider analysts or dispatchers
to be in charge of rebalancing operations.

\begin{figure}

{\centering \includegraphics{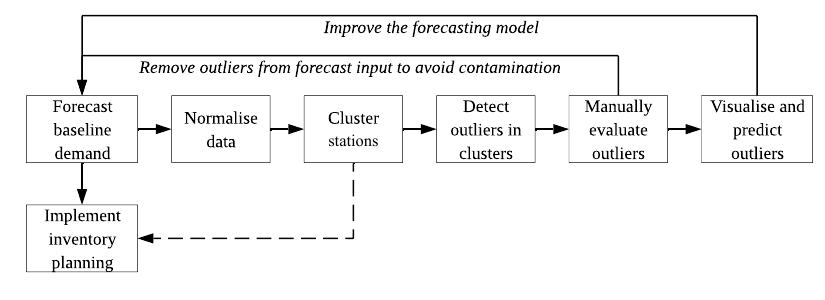}

}

\caption{\label{fig-process_map}Flowchart of process for analysing
bike-sharing demand data. Figure adapted from \citet{Rennie_thesis}.}

\end{figure}

To combine automated outlier detection, manual analysis, demand
forecasts, and planning, we suggest the following process for analysing
bike-sharing demand data (see Figure~\ref{fig-process_map}): First, a
baseline demand forecast supports anticipative planning, e.g.~of
inventory levels. Second, this baseline can be used to normalise
observed usage data. Using the resulting observations, analysts can
cluster stations with similar usage patterns to support both planning
adjustments and outlier detection. When detecting outliers in a
cluster's usage patterns, these are visualised to enable manual outlier
evaluation. Insights from this analysis can be used to both clean the
data that underlies the baseline forecast and to extend the baseline
forecasting model.

In this paper, we analyse the Capital Bikeshare data set, which is
publicly available at \citet{CaBi}. This data set is commonly used to
test forecasting approaches for bike-sharing
\citep[\citet{Hamilton2018}]{Ma2015}, yet these methods typically do not
account for outliers. In Section \ref{sec-data}, we introduce the data
set and perform an exploratory analysis. Section \ref{sec-st_patterns}
and Section \ref{sec-clustering_ch3} then model the temporal and spatial
patterns in demand for bike-sharing. In Section
\ref{sec-outliers_method}, we provide a methodology for identifying
outlying demand for bike-sharing services. The results of applying the
outlier detection method to the Capital Bikeshare data are then
discussed in Section \ref{sec-discussion}.

In summary, this paper contributes (i) an in-depth analysis of temporal
patterns in usage of Capital Bikeshare services; (ii) a method for
spatial clustering of bike-sharing stations based on geographic
proximity and similarity of usage patterns; (iii) an investigation of
temporal trends in detected outliers and the factors that may cause
them; and (iv) an analysis of spatial patterns of the outliers detected.
Our methodology is data-driven and general by design, and not tailored
to specifics related to Washington D.C., and can thus be readily applied
to all bike-sharing data sets around the world.

\hypertarget{sec-data}{%
\section{Capital Bikeshare data}\label{sec-data}}

The Capital Bikeshare data set spans a three year period, from January 1
2017 to December 31 2019. It describes every recorded trip by its time
of pick-up, time of drop-off, pick-up station location, and drop-off
station location. The data set features only those 578 stations that
recorded at least one pick-up or drop-off within the recorded time
frame.

\begin{figure}

{\centering \includegraphics{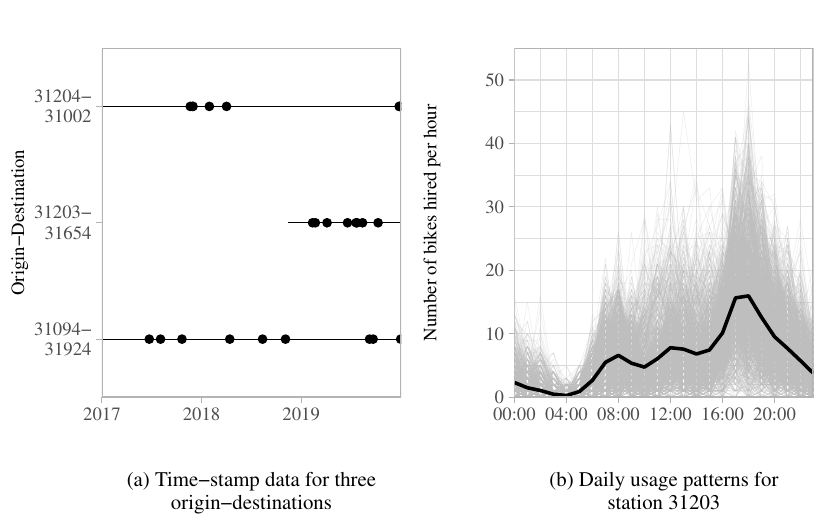}

}

\caption{\label{fig-bookings_od_terminal}Origin-destination (O-D) level
data and aggregated daily usage patterns, with mean usage pattern
indicated in (b). Figure adapted from \citet{Rennie_thesis}.}

\end{figure}

Out of a potential 334,084 unique origin-destination (O-D) pairs, the
data set records 105,735 O-D pairs that customers completed based on
pick-ups and drop-offs. As examples, the times of bike rentals for three
different O-D pairs are shown in
Figure~\ref{fig-bookings_od_terminal}(a), with each dot representing one
journey. Note that station 31654 opened in November 2018, and so data is
only available from that date onward for O-D pair 31203-31654.

\hypertarget{data-cleansing}{%
\paragraph{Data cleansing}\label{data-cleansing}}

If the first use of a station in the data set was not January
1\textsuperscript{st} 2017, we check historical data from 2016 for any
usage to determine if the station was open. If there are no earlier
bookings, we consider the station as newly opened from the time of its
first recorded trip. Capital Bikeshare pre-process the data to remove
trips that are made by staff for system maintenance and any trips with a
journey time of less than 60 seconds (as these may be false starts).

\hypertarget{data-aggregation}{%
\paragraph{Data aggregation}\label{data-aggregation}}

Given the large number of O-D pairs, very few journeys are recorded per
unique pair on average. This makes it difficult to detect meaningful
patterns, or any deviation from such a pattern, on the O-D level. When
numbers are this small, noise dominates over any trend, as also pointed
out in related research on forecasting slow-moving retail products
\citep{Jha2015}. To alleviate the problem of small numbers, we aggregate
trips as pick-up and drop-off events, considering usage per station
rather than O-D pairs. To further reduce the problem of sparse
observations, and to make observations comparable over time, we
aggregate usage by hour of day {[}\citet{Petropoulos2015}. Specifically,
we define the daily usage pattern to be a time series of the number of
times per hour that a station is used, either pick-up or drop-off -- see
Figure~\ref{fig-bookings_od_terminal}(b). When considering pick-ups and
drop-offs separately, we differentiate the daily pick-up pattern and the
daily drop-off pattern.

\begin{figure}

{\centering \includegraphics{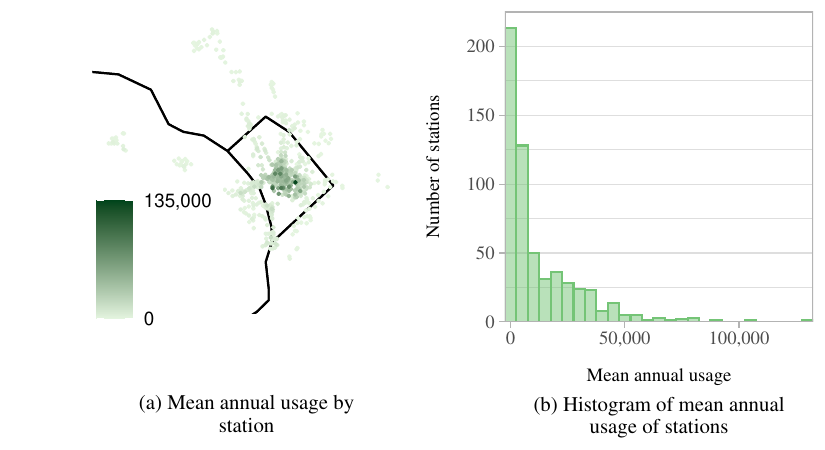}

}

\caption{\label{fig-usage_terminal}Mean annual usage per station with
(a) indicating higher usage nearer the city centre, and (b) showing a
histogram of number of stations under varying levels of usage, in
intervals of 5,000. Figure adapted from \citet{Rennie_thesis}.}

\end{figure}

\hypertarget{exploratory-analysis-spatial-variation.}{%
\paragraph{Exploratory analysis: Spatial
variation.}\label{exploratory-analysis-spatial-variation.}}

The total usage varies greatly across stations, with those closer to the
centre of Washington D.C. being more popular on average.
Figure~\ref{fig-usage_terminal}(a) visualises this idea by indicating
the mean annual usage per station across the region. The most popular
stations observe more than 130,000 uses per year, whereas the least
popular stations observe less than one on average. Over half of the
stations (51\%) recorded fewer than 5,000 pick-up or drop-off events per
year. To indicate the distribution, Figure~\ref{fig-usage_terminal}(b)
provides the mean annual usage per station in a histogram.

\hypertarget{exploratory-analysis-temporal-variation.}{%
\paragraph{Exploratory analysis: Temporal
variation.}\label{exploratory-analysis-temporal-variation.}}

\begin{figure}

{\centering \includegraphics{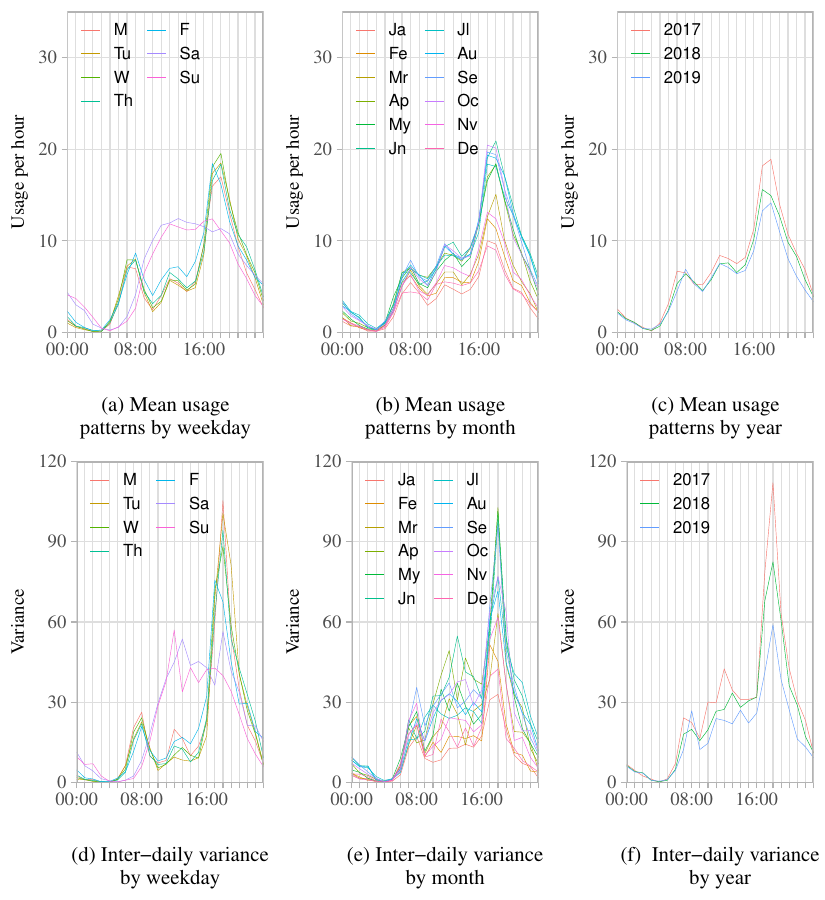}

}

\caption{\label{fig-mean_var}Mean usage patterns and inter-daily
variance for station 31203 by hour of day, which is a representative
pattern as seen across the network. Figure originally included in
\citet{Rennie_thesis}.}

\end{figure}

In addition to daily usage patterns varying across space, there is also
significant temporal variation. Figure~\ref{fig-mean_var} that there are
significantly different mean usage patterns and inter-daily variance for
different days, months, and, to some extent, years. Here, we define
inter-daily variance as the daily variability in the usage at a station
at a given hour of the day. We present results here for only a single
station as these mean and variance patterns can vary across stations,
though similar differences between, e.g.~days are observed across most
stations. This variation in patterns across stations motivates the
functional regression model described in Section \ref{sec-func_reg}.

\hypertarget{sec-st_patterns}{%
\section{Modelling baseline temporal usage
patterns}\label{sec-st_patterns}}

As discussed in Section \ref{sec-data}, usage patterns vary across time
and space. If we do not first remove temporal patterns observed in
baseline demand, any outlier detection procedure will likely simply
detect baseline trend characteristics as outliers. For example, there is
a much higher level of variability in demand on weekends in summer. If
we failed to account for this before performing the outlier detection
procedure, many of the detected outliers would occur on Saturdays in the
summer months. By first accounting for known temporal patterns, the
detected outliers are more likely to be genuine outliers rather than
explainable patterns already known to analysts.

Similarly, if we do not account for spatial baseline variability and
instead aggregate data across all stations then we will simply detect
unused or extremely busy stations as outliers. Conversely, if we assume
all stations behave independently, then the increased noise makes it
more difficult to detect outlying usage patterns. As such, before
implementing the outlier detection procedure outlined later in Section
\ref{sec-outliers_method}, we carry out a two-step process to (i) remove
known temporal patterns; and (ii) spatially cluster stations which
behave similarly. These two steps are key in identifying meaningful
outliers, as we shall show.

\hypertarget{background-bike-sharing-demand-forecasting}{%
\subsection{Background: Bike-sharing demand
forecasting}\label{background-bike-sharing-demand-forecasting}}

Within the bike-sharing literature, a range of techniques have been
considered to predict demand, both spatially and temporally.
\citet{Zhou2018} apply a Markov Chain based model to predict daily
pick-ups and drop-offs at each station within the Zhongshan City
bike-sharing system. The problem of predicting demand in the presence of
spatial heterogeneity is further considered by \citet{Gao2021} who
estimate a distance decay function and then use multiple linear
regression to predict temporal demand in the dockless bike-sharing
system in Shanghai. Dockless bike-sharing systems are also discussed by
\citet{Xu2018}, who use long short-term memory neural networks to
predict demand, and capture the spatial and temporal imbalance in usage.
\citet{Sohrabi2021} use a combination of pattern recognition on historic
data traffic patterns and \(K\)-nearest neighbours to make
spatiotemporal demand predictions over short time horizons (between 15
minutes and 4 hours), for the Capital Bikeshare data.

\begin{figure}

{\centering \includegraphics{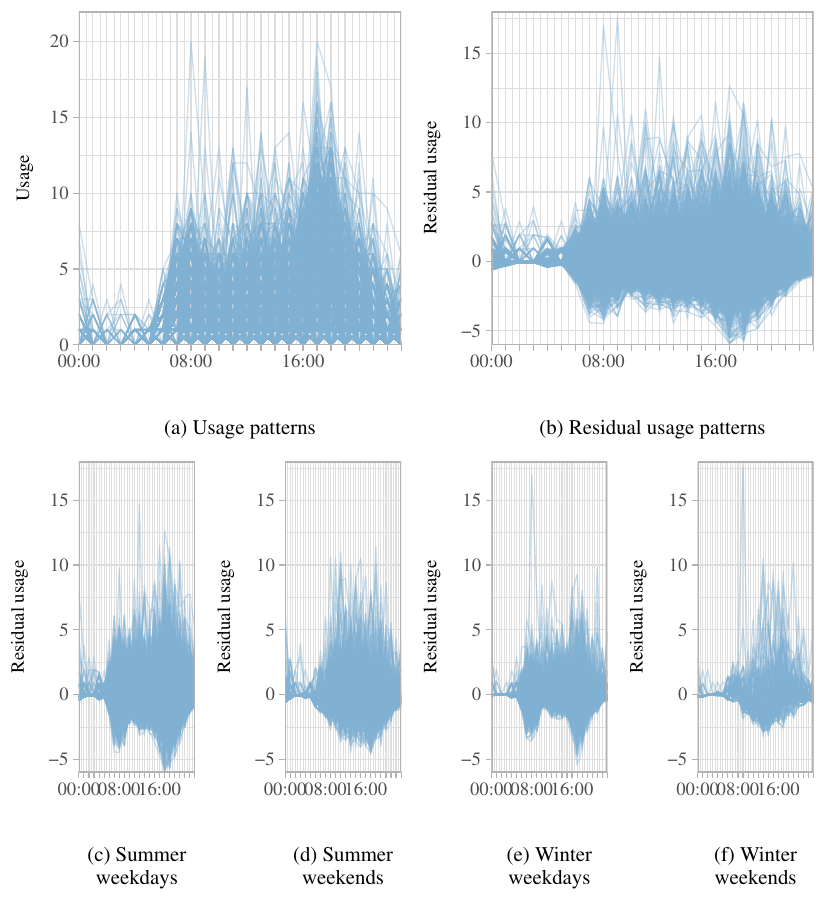}

}

\caption{\label{fig-residuals}Residual usage patterns for station 31005.
Figure originally included in \citet{Rennie_thesis}.}

\end{figure}

The choice of forecasting approach will likely affect the outcome of
outlier detection. In the following, we discuss and apply two methods of
predicting the baseline temporal patterns in the data: (i) functional
regression to account for changes in mean; and (ii) temporal
partitioning to account for changes in variance.

Beyond the model presented here, alternative approaches could be used to
account for trend and seasonality, and establishing a baseline for
bike-sharing demand. In general, any forecasting or modelling approach
from which residuals can be obtained could be used instead. After the
temporal patterns have been accounted for and the residuals obtained, we
are then able to analyse the spatial correlations to group together
stations which deviate from the baseline demand forecast in a similar
way, as we shall discuss later in Section \ref{sec-clustering_ch3}.

\hypertarget{sec-func_reg}{%
\subsection{Functional regression}\label{sec-func_reg}}

Mean daily usage patterns differ systematically across days of the week,
months, and years. We apply a functional regression model
\citep{Ramsay2009} to remove the different mean patterns. We will
demonstrate this process on the daily usage patterns of station 31005 as
shown in Figure~\ref{fig-residuals}(a).

Let \(x_{ns}(t)\) be the usage pattern for day \(n\) for station \(s\).
We implement the following functional regression model:

\begin{equation}\protect\hypertarget{eq-funcreg}{}{
\begin{split}
  x_{n,s}(t) = \beta_{0,s}(t) +
  \textcolor{cyan}{\beta_{1,s}(t)\mathds{1}_{Mon_{n}} + \beta_{2,s}(t)\mathds{1}_{Tue_{n}} + \beta_{3,s}(t)\mathds{1}_{Wed_{n}} +} \\ \underbrace{\textcolor{cyan}{\beta_{4,s}(t)\mathds{1}_{Thu_{n}} + \beta_{5,s}(t)\mathds{1}_{Fri_{n}} + \beta_{6,s}(t)\mathds{1}_{Sat_{n}}+}}_\textrm{\textcolor{cyan}{Day}} \\ 
  \textcolor{purple}{\beta_{7,s}(t)\mathds{1}_{Jan_{n}} + \beta_{8,s}(t)\mathds{1}_{Feb_{n}} + \beta_{9,s}(t)\mathds{1}_{Mar_{n}} +} \\
  \textcolor{purple}{\beta_{10,s}(t)\mathds{1}_{Apr_{n}} + \beta_{11,s}(t)\mathds{1}_{May_{n}} + \beta_{12,s}(t)\mathds{1}_{Jun_{n}}+}
  \\ 
  \textcolor{purple}{\beta_{13,s}(t)\mathds{1}_{Jul_{n}}+\beta_{14,s}(t)\mathds{1}_{Aug_{nl}}+\beta_{15,s}(t)\mathds{1}_{Sep_{n}}+}
  \\ 
  \underbrace{\textcolor{purple}{
  \beta_{16,s}(t)\mathds{1}_{Oct_{n}}+\beta_{17,s}(t)\mathds{1}_{Nov_{n}}+}}_\textrm{\textcolor{purple}{Month}} \\ 
 \underbrace{\textcolor{blue}{\beta_{18,s}(t)\mathds{1}_{2017_{n}} + \beta_{19,s}(t)\mathds{1}_{2018_{n}}}}_\textrm{\textcolor{blue}{Year}} + e_{n,s}(t).
\end{split}
}\label{eq-funcreg}\end{equation}

where e.g., \(\mathds{1}_{Mon_{n}} =1\) if day \(n\) relates to a
Monday, \(0\) otherwise. Here, \(\beta_{0,s}(t)\) represents the mean
usage pattern for a Sunday in December 2019. Appendix
\ref{sec-app-model_comp} contains details of the model selection process
where we consider the significance of each of the factors (day, month,
year) for a range of stations. The vast majority of stations select the
full model containing all three factors as the best-fitting model.

As Figure~\ref{fig-residuals}(b) indicates, the core of the distribution
of residuals is symmetric around 0 as desired. The majority of
``spikes'\,' in usage are caused by increased demand, resulting in a
slight positive skew to the residual patterns. We note that the variance
of these residuals is clearly not constant over time and we shall
discuss this shortly. Further discussion of the residual distribution is
included in Appendix \ref{sec-app-ridges}. Other features of the usage
patterns including positive skew, and inter-daily correlation are
discussed in Appendices \ref{sec-app-skew} and \ref{sec-app-acf}.

\hypertarget{temporal-partitioning.}{%
\subsection{Temporal partitioning.}\label{temporal-partitioning.}}

Our functional regression approach accounts for different mean usage
patterns, but it does not account for the differing inter-daily
variances. The simplest option to obtain a data set with homogeneous
inter-daily variance is to temporally partition the full data set. While
we could partition based on each weekday, month, and year, this would
result in around 4 observations per partition - an insufficient number
to inform outlier detection. In deciding how to partition the data,
there is a trade-off between having reasonably constant inter-daily
variance within each group and ensuring there is enough data within each
group in order to establish patterns. Therefore, we group together days,
months, and years where the inter-daily variances are sufficiently
similar.

From Figure~\ref{fig-mean_var}, it is clear that weekdays (Mon-Fri) are
similar to each other, and weekends (Sat-Sun) are also similar to each
other. The differences between the inter-daily variance across different
months is less clear. Defining \textbf{summer} as April through to
October, then months within summer exhibit similar inter-daily variance
patterns, as do months within winter (November to March). Further
analysis of the variance in Appendix \ref{sec-app-preproc} supports this
partitioning. All years are grouped together. This results in four
partitions: (i) summer weekdays, (ii) winter weekdays, (iii) summer
weekends, and (iv) winter weekends, as displayed in
Figure~\ref{fig-residuals} c--f.~Note that we do not attempt to remove
the \textbf{intra}-daily variability of these residuals with further
parametric modelling, as instead we turn to functional data analysis to
detect outlying curves from these residual daily usage patterns.

The choice of partition is important and should reflect the choices made
in the planning process, e.g.~with regard to inventory redistribution.
If the increased inter-daily variance on weekends, for example, is
already known and accounted for in planning, such that there are
different schedules for redistribution, then partitioning as we propose
would be appropriate. However, if the general planning process
(including demand forecast and inventory optimisation) assumes
uniformity across all days of the week, it would then be informative to
do the same in the outlier detection to flag the weekend effect when it
occurs.

\hypertarget{sec-clustering_ch3}{%
\section{Clustering stations by spatial usage
patterns}\label{sec-clustering_ch3}}

When outlier demand is driven by factors such as regional events or
weather, we expect it to affect more than a single, isolated station. At
the same time, we cannot assume that all stations experience outlier
demand at the same time and in the same way. Therefore, we first cluster
the stations such that those in the same cluster are likely to
experience similar effects from demand outliers.

We propose a two-stage process to determine which stations should be
clustered. First, we construct a graph based on the geographic
co-ordinates of the stations to determine which stations are permitted
to be in the same cluster based on geographic distance. Secondly, we
follow an idea from \citet{Zahn1971} who suggests the removal of edges
from a graph's minimum spanning tree (MST) as a method of finding
clusters of nodes.

The first step of constructing the graph is non-trivial. Graph
construction in the bike-sharing setting is more open-ended than in
situations where mobility networks rely on established legs, as, e.g.,
in the railway application studied in \citet{Rennie_thesis}. For
bike-sharing, direct journeys between any two stations are possible, so
that in theory, all stations could be vertices in a fully connected
graph. Although we could simply add edges between every node i.e.~a
complete graph, there are two reasons for not doing so: (i) For the
purposes of aiding planning, we do not want two stations which are
geographically far apart to be in the same cluster if no stations in
between are similar to both. (ii) The algorithm used to compute the MST
is slowed down by an increased number of edges, and due to the greedy
nature of it, we are more likely to end up at a non-optimal solution if
we add in extraneous edges.

\hypertarget{graph-construction-from-geographical-distance.}{%
\subsection{Graph construction from geographical
distance.}\label{graph-construction-from-geographical-distance.}}

We first construct a graph where the nodes represent the stations and
the edges indicate which stations are permitted to be in the same
cluster. This approach implicitly assumes that similarity of usage is
driven by the stations' geographical proximity. That is, if two stations
are close together, potential customers are more likely to treat them as
interchangeable, causing similar usage patterns.

In the Capital Bikeshare data set, stations are more densely distributed
in the centre of D.C., so that customers can choose from a large variety
of stations. We expect this to render them more sensitive to distance,
such that they are less willing to travel to a more distant station.
Therefore, we use different criteria to add an edge between stations
depending on how close to the centre of D.C. those stations are.

\begin{figure}

{\centering \includegraphics{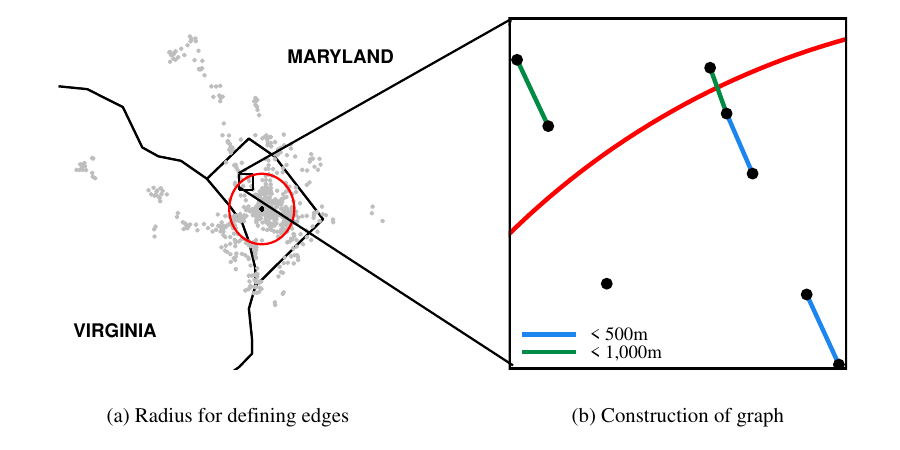}

}

\caption{\label{fig-mst-two}Graph construction when \(R = 5000m\),
\(D_{inner} = 500m\), and \(D_{outer} = 1,000m\). Figure originally
included in \citet{Rennie_thesis}.}

\end{figure}

To identify the dense city-centre, we establish a circle around the
centre of D.C. of radius \(R\), with the median co-ordinates of all
stations as the centre, as shown in Figure~\ref{fig-mst-two}(a) with
\(R = 5000m\). We add an edge between stations \(i\) and \(j\) if: (i)
both stations lie inside the radius, and are less than \(D_{inner}\)
apart; or (ii) one or both stations lie outside the radius \(R\), and
are less than \(D_{outer}\) apart.

Not all stations that are geographically close exhibit similar usage
patterns e.g.~due to proximity to railway stations. Therefore, to
quantify how similar the usage patterns of two stations are, we add
weights to the edges of the graph. For each edge between stations \(i\)
and \(j\), we also compute an edge weight representing the dissimilarity
between the usage patterns for those stations. The edge weights are
given by:

\begin{equation}
    w{(i, j)} = 1 - \rho(i, j),
\end{equation}

where \(\rho(i, j)\) is the average functional dynamical correlation
\citep{Dubin2005} between the daily usage patterns for stations \(i\)
and \(j\). Here, the average correlation is based on the correlations
between daily usage patterns across the entire time period considered
(2017 - 2019), as there is no evidence of the clusters changing over
time. However, if the correlations (and therefore clusters) are changing
over time, a moving window approach could be used to update the average
correlation and clusters over time.

\hypertarget{minimum-spanning-tree-clustering.}{%
\subsection{Minimum spanning tree
clustering.}\label{minimum-spanning-tree-clustering.}}

We apply a minimum spanning tree approach to cluster stations that are
connected in the geographical proximity graph. A graph's spanning tree
is a subgraph that includes all vertices in the original graph and a
minimum number of edges, such that the spanning tree is connected. If
the original graph is disconnected, we compute a spanning tree for each
component -- termed a \textbf{spanning forest}. A minimum spanning tree
(MST) is the spanning tree with the minimum sum of edge weights. Since
the graph is weighted, we use Prim's algorithm \citep{Prim1957} to
calculate the MST.

To obtain the clusters from the MST, we set a threshold,
\(\rho_{\tau}\), for the correlation and remove all edges with weights
above \(1 - \rho_{\tau}\).

\hypertarget{sec-clust_output}{%
\subsection{Clustering results: daily usage
patterns}\label{sec-clust_output}}

\begin{figure}

{\centering \includegraphics{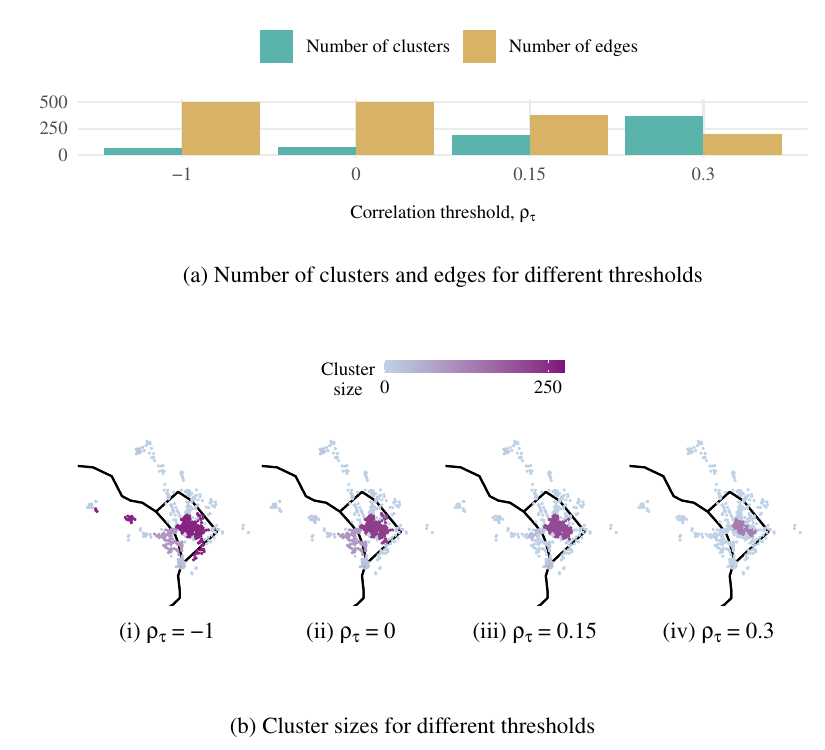}

}

\caption{\label{fig-clust_output}Clustering of stations under different
values of \(\rho_{\tau}\). Figure adapted from \citet{Rennie_thesis}.}

\end{figure}

Figure~\ref{fig-clust_output} visualises the outcome from four different
values of \(\rho_{\tau}\). These values of \(\rho_{\tau}\) are chosen to
illustrate the clustering for two reasons: (i) \(\rho_{\tau}=-1\)
indicates that all initially connected edges stay in place. In fact, the
minimum correlation observed is -0.57, and any threshold between -1 and
-0.57 results in all edges of the MST remaining in place. (ii) 90\% of
the observed correlations lie between 0 and 0.3, therefore the values of
\(\rho_{\tau}\) = 0, 0.15, and 0.3 demonstrate the clustering when the
threshold is close to the minimum, mean, and maximum correlations.

Figure~\ref{fig-clust_output} can be used by analysts to determine the
most appropriate threshold, depending on which aspect of planning they
are considering. Figure~\ref{fig-clust_output}(a) shows how the number
of clusters (and edges) changes with the choice of clustering threshold
- with little difference seen between a threshold of -1 and 0. By
inspecting which stations are clustered together, if an analyst has
expert knowledge regarding which stations are likely to behave
similarly, they can cross-check with the clustering and choose the
threshold which supports this decision. Figure~\ref{fig-clust_output}(b)
visualises the sizes the clusters that each station belongs to,
demonstrating the non-uniform distribution of cluster size across the
geographic area. This can also be used to determine an appropriate
threshold. For example, if an analyst is interested in the general
demand patterns of central D.C., they can choose a threshold that
highlights all of central D.C. in a single large cluster
e.g.~\(\rho_{\tau}\)=0. In contrast, if the analyst is more interested
in obtaining clusters of similar size,
Figure~\ref{fig-clust_output}(b-iv) would guide them towards a higher
threshold.

Across all thresholds, the stations closer to the centre of D.C. form a
larger cluster, with those further away from the centre branching into
smaller clusters. Clearly, the choice of threshold values
\(\rho_{\tau}\) impacts the precise clustering results.

The distance parameters, \(\{R,D_{inner},D_{outer}\}\), also affect
clustering. The number of clusters increases as \(\rho_{\tau}\) or \(R\)
is increased, whereas increasing \(D_{inner}\) or \(D_{outer}\) has the
opposite effect. There is an inverse relationship between the number of
clusters and the uniformity of cluster size. As the number of clusters
increases, individual stations tend to split off to form their own
cluster whilst the majority of stations remain in the large central
cluster, resulting in decreased uniformity of cluster size. For
decision-making, clusters of similar sizes are often more informative
(compared to a large cluster consisting of most stations, and the
remaining stations each in their own cluster).

We leave the choice of parameters to analyst input, such that analysts
may use their expertise to select appropriate values based on the
visualisation and their business case \citep{Vock2021}. For the
remainder of this paper, unless otherwise specified, we set the
parameter values as \(\rho_{\tau}\) = 0.15, \(R\) = 5000m, \(D_{inner}\)
= 500m, and \(D_{outer}\) = 1000m. These values are chosen to balance
the number of clusters with more similar cluster sizes. Appendix
\ref{sec-app-cluster_params} includes further details on the reasoning
for these choices.

\hypertarget{clustering-results-daily-pick-up-and-drop-off-patterns}{%
\subsection{Clustering results: daily pick-up and drop-off
patterns}\label{clustering-results-daily-pick-up-and-drop-off-patterns}}

\begin{figure}

{\centering \includegraphics{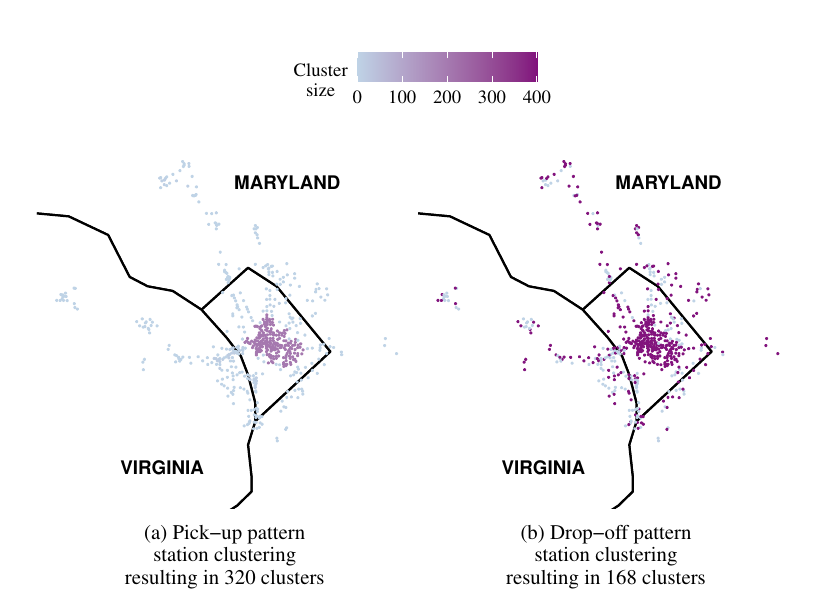}

}

\caption{\label{fig-start_end}Comparison of clustering stations based on
pick-up and drop-off patterns for \(\rho_{\tau}\)=0.15, \(R\) = 5000m,
\(D_{inner}\) = 500m, and \(D_{outer}\) = 1000m. Figure adapted from
\citet{Rennie_thesis}.}

\end{figure}

So far, we have focused on clustering stations based on the similarity
of their daily usage patterns. However, when considering forecasting for
inventory rebalancing, differentiating pick-ups and drop-offs is highly
important. Depending on the aggregation level of forecasting, it may be
desirable to consider separate clusterings for drop-off and pick-up
patterns. Figure~\ref{fig-start_end} shows that drop-off patterns tend
to be more homogeneous, in comparison to pick-up patterns, resulting in
fewer clusters for the same correlation threshold.

\begin{figure}

{\centering \includegraphics{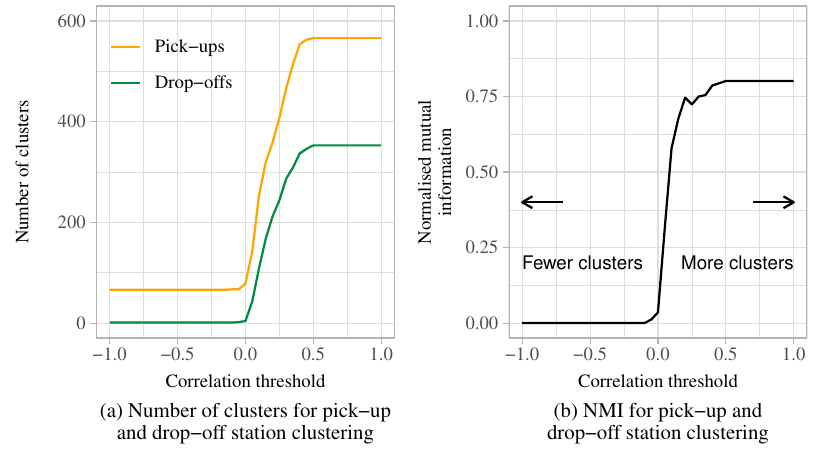}

}

\caption{\label{fig-start_end_nmi}Comparison of pick-up and drop-off
station clustering. Figure adapted from \citet{Rennie_thesis}.}

\end{figure}

Figure~\ref{fig-start_end_nmi}(a) shows that this increased homogeneity
of drop-off patterns is consistent across all values of the correlation
threshold, \(\rho_{\tau}\). Although the number of clusters resulting
from both pick-up and drop-off station clustering follows a similar
relationship with the correlation threshold -- increasing steeply
between 0 and 0.4 -- the drop-off clustering consistently results in
fewer clusters.

To formally compare the output of these two clusterings, we use the
normalised mutual information (NMI) \citep{Amelio2015}. The NMI is 1 if
two clusterings are identical, and 0 if they are completely different
(see Appendix \ref{sec-app-nmi_def_ch3} for details).
Figure~\ref{fig-start_end_nmi}(b) shows that the similarity of the
pick-up and drop-off clusterings are highly dependent on the correlation
threshold. When a low threshold is used, the clusterings are completely
different. However, as the correlation threshold increases above 0.25,
the clusterings become more similar, achieving an NMI of around 0.75.

This evidence that pick-ups and drop-offs are not spatially homogeneous
motivates the need for separate forecasting of the two. The differences
across the varying threshold also indicates that the need for separate
forecasting is more critical when considering a larger area i.e.~when
considering total demand, but is less critical over smaller areas closer
to the station level. Monitoring the difference in the number of
clusters and the similarity of the two clusterings can help analysts to
decide on the level of forecasting. Analysts could also examine changes
in the NMI over time for a given correlation threshold. For example, if
the pick-up and drop-off clusterings are becoming more similar to each
other over time, this could indicate increasing levels of homogeneity in
pick-up patterns.

\hypertarget{sec-outliers_method}{%
\section{Detecting outliers within a cluster of
stations}\label{sec-outliers_method}}

\begin{figure}

{\centering \includegraphics{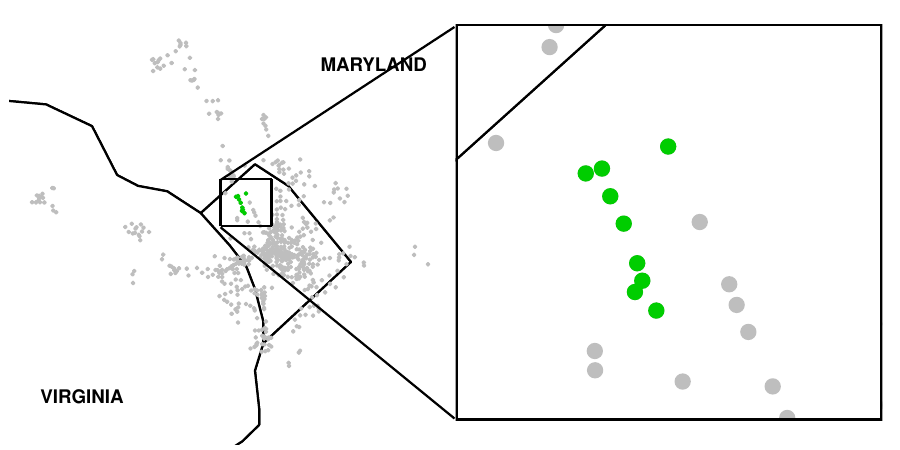}

}

\caption{\label{fig-mst_zoom}Cluster chosen for further investigation.
Figure originally included in \citet{Rennie_thesis}.}

\end{figure}

To demonstrate the outlier detection procedure, we focus on one of the
resulting clusters. The nine stations in the cluster we consider are
highlighted in green in Figure~\ref{fig-mst_zoom}.

Figure~\ref{fig-mst_zoom} demonstrates how the currently analysed
cluster may be highlighted for analysts. On the one hand, the location
of the cluster within the D.C. area can provide contextual information
for analysts in the search for an explanation of outlier demand. On the
other hand, the zoomed in section on the right shows how the stations
relate to one another within the cluster. This could be useful if the
outlier demand is not detected in all stations. For example, all but one
of the green cluster stations lie in a relatively straight line. If the
station which lies to the North East of the main group of stations in
the cluster behaves differently, analysts can look to nearby clusters
for further information.

To identify outlier demand in usage patterns, we use the notion of
\emph{statistical depth}. In statistics, depth provides an ordering of
observations, where those near the centre of the distribution have
higher depth and those far from the centre have lower depth. In the case
where each observation is a time series of usage throughout the day, the
\emph{functional depth} can measure how close to the central trajectory,
i.e.~median usage pattern, each daily usage pattern is. Therefore, to
measure the outlyingness of each daily usage pattern, we calculate its
functional depth (with respect to other daily usage patterns that lie in
the same partition of data). Days whose usage pattern has lower
functional depth are more outlying. In particular, if the depth is below
some threshold, we classify the day as an outlier.

For each partition of data, \(p\), and for each station \(s\), we
calculate a threshold, \(C_{s,p}\), for the functional depth as per
\citet{Febrero2008}. To calculate the threshold, we (i) resample the
daily usage patterns with probability proportional to their functional
depths (such that any usage patterns affected by outlier demand are less
likely to be resampled), (ii) smooth the resampled patterns, and (iii)
sets the threshold \(C_{s,p}\) as the median of the \(1^{st}\)
percentile of the functional depths of the resampled patterns.

\begin{figure}

{\centering \includegraphics{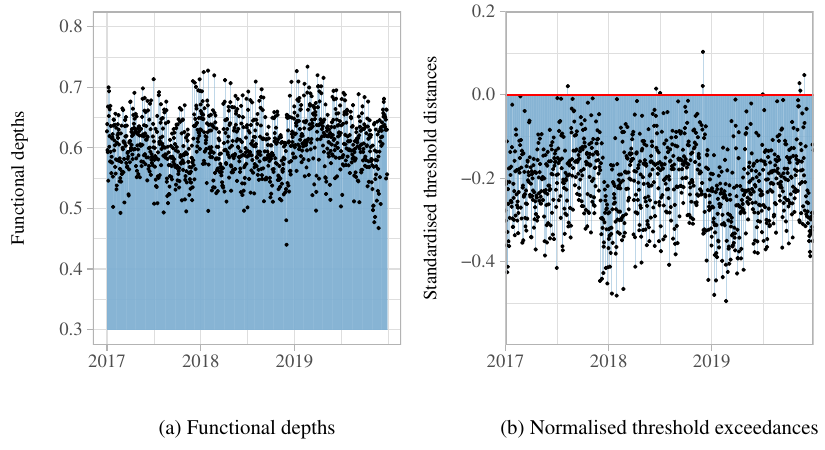}

}

\caption{\label{fig-diffs}Normalisation of the functional depths,
exemplified for station 31316 where there are two partitions of data
(summer/winter). Figure originally included in \citet{Rennie_thesis}.}

\end{figure}

Let \(d_{n,s,p}\) be the functional depth for day \(n\) (which occurs in
partition \(p\)) for station \(s\). We then transform the functional
depths to lie between 0 and 1 such that they are comparable between
different stations and aggregated over the different partitions of data.
Define \(z_{n,s}\) to be the normalised functional depth on day \(n\)
for station \(s\):

\begin{equation}
    z_{n,s} = \sum_{p=1}^P \left( \mathds{1}_{n \in p}\left( \frac{C_{s,p} - d_{n,s,p}}{C_{s,p}}\right) \right).
\end{equation}

The functional depths for station 31303 are shown in
Figure~\ref{fig-diffs}(a), and their normalised counterparts in
Figure~\ref{fig-diffs}(b). Figure~\ref{fig-diffs}(a) provides a way to
check for unaccounted for trend and seasonality in the usage patterns.
However, much like univariate regression residuals which can be used to
visually identify residuals patterns, the functional depths should
appear random with no obvious patterns. If an analyst can identify a
pattern in the functional depths, this would suggest that the
forecasting model may need to be reconsidered. Weekdays could be
highlighted in different colours to help identify temporal patterns on a
smaller scale. Figure~\ref{fig-diffs}(b) can also be used by analysts to
check how many non-outlier days are close to, but do not exceed, the
threshold. Analysts can use this information to manually vary the
threshold to detect further outliers they perceive to be false
negatives.

\begin{figure}

{\centering \includegraphics{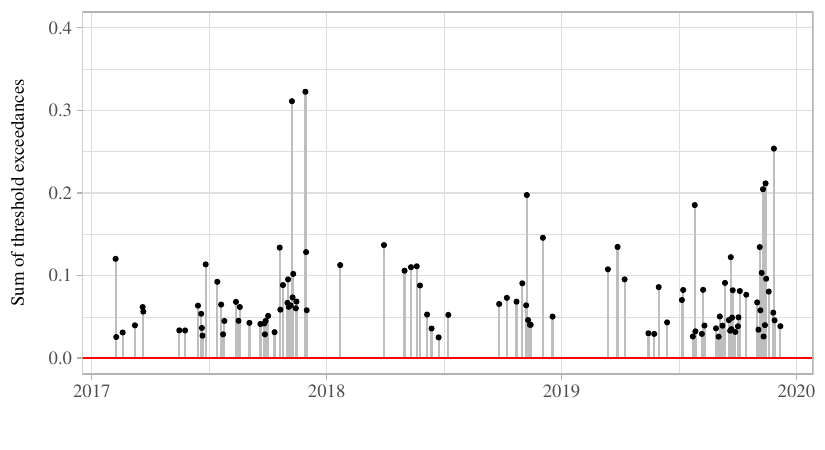}

}

\caption{\label{fig-zn}Sum of threshold exceedances, \(z_n\). Figure
originally included in \citet{Rennie_thesis}.}

\end{figure}

We then define the sum of threshold exceedances across all \(S\)
stations in the cluster to be:

\begin{equation} 
 z_n = \sum_{s=1}^{S} z_{n,s} \mathds{1}_{\{z_{ns} > 0\}}.
\end{equation}

The values of \(z_{n}\) for this cluster are shown in
Figure~\ref{fig-zn}. The value of \(z_n\) is only positive for days
which have been classified as an outlier.

\hypertarget{computing-outlier-severity}{%
\subsection{Computing outlier
severity}\label{computing-outlier-severity}}

Although the values of \(z_n\) give an indication of how severe the
outlier is (with \(z_n\) being larger if the magnitude of the outlier
demand is larger, or if it affects a larger number of stations), we wish
to make the severity easier to interpret across different clusters.
Therefore, we fit a distribution to the sum of threshold exceedances and
use the non-exceedance probability given by the CDF of the distribution
as a measure of severity.

\begin{figure}

{\centering \includegraphics{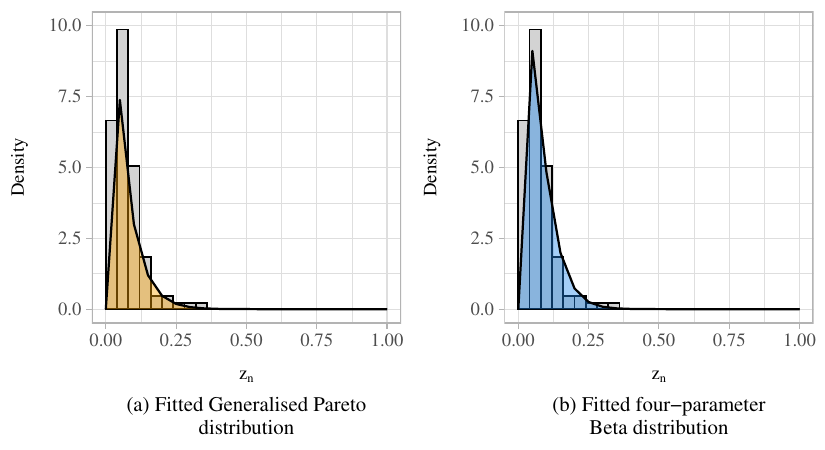}

}

\caption{\label{fig-gpd_fit}Comparison of fitted distributions. Figure
originally included in \citet{Rennie_thesis}.}

\end{figure}

In contrast to \citet{Rennie2021b} who fit a generalised Pareto
distribution (GPD) to the sum of threshold exceedances, here we fit a
four-parameter Beta distribution \citep{Carpenter2001}. For a GPD,
assuming the shape parameter is non-negative, the support has no upper
bound. In this application the upper bound is finite and known to be
equal to the number of stations within the cluster. Since \(z_{n,s}\)
lies between 0 and 1, the sum across \(S\) stations must lie between 0
and \(S\). Therefore, a four parameter Beta distribution, bounded on
\((0,S)\) is likely to provide a better fit -- see
Figure~\ref{fig-gpd_fit}.

The severity of an outlier on day \(n\), \(\theta_n\), is therefore
given by the CDF of the four parameter Beta distribution:

\begin{equation} 
    \theta_n = F_{(\alpha, \beta, a, c)}(z_n) = 
    \int_{0}^{z_n} \frac{(q)^{\alpha -1} (S - q)^{\beta -1}}{B(\alpha, \beta)S^{\alpha + \beta + 1}} dq,
\end{equation}

where \(B(\alpha, \beta)\) is the Beta function. This results in an
outlier severity, between 0 and 1, for each cluster on each day, as
exemplified in Table~\ref{tbl-severity}.

We differentiate \emph{positive} and \emph{negative} outliers. Positive
outliers are primarily caused by increased usage i.e.~where the sum of
the functional residual is greater than zero, whereas negative outliers
indicate a shortfall in usage.

\begin{table}

\end{table}

\hypertarget{tbl-severity}{}
\begin{longtable}[]{@{}
  >{\centering\arraybackslash}p{(\columnwidth - 8\tabcolsep) * \real{0.1690}}
  >{\centering\arraybackslash}p{(\columnwidth - 8\tabcolsep) * \real{0.2817}}
  >{\centering\arraybackslash}p{(\columnwidth - 8\tabcolsep) * \real{0.2535}}
  >{\centering\arraybackslash}p{(\columnwidth - 8\tabcolsep) * \real{0.1549}}
  >{\centering\arraybackslash}p{(\columnwidth - 8\tabcolsep) * \real{0.1408}}@{}}
\toprule()
\begin{minipage}[b]{\linewidth}\centering
Date
\end{minipage} & \begin{minipage}[b]{\linewidth}\centering
Cluster 1
\end{minipage} & \begin{minipage}[b]{\linewidth}\centering
Cluster 2
\end{minipage} & \begin{minipage}[b]{\linewidth}\centering
Cluster 3
\end{minipage} & \begin{minipage}[b]{\linewidth}\centering
\ldots{}
\end{minipage} \\
\midrule()
\endfirsthead
\toprule()
\begin{minipage}[b]{\linewidth}\centering
Date
\end{minipage} & \begin{minipage}[b]{\linewidth}\centering
Cluster 1
\end{minipage} & \begin{minipage}[b]{\linewidth}\centering
Cluster 2
\end{minipage} & \begin{minipage}[b]{\linewidth}\centering
Cluster 3
\end{minipage} & \begin{minipage}[b]{\linewidth}\centering
\ldots{}
\end{minipage} \\
\midrule()
\endhead
11/01/2017 & 0.033 \(\downarrow\) & - & - & \(\hdots\) \\
12/01/2017 & 0.852 \(\uparrow\) & 0.720 \(\uparrow\) & - & \(\hdots\) \\
\(\vdots\) & \(\vdots\) & \(\vdots\) & \(\vdots\) & \(\ddots\) \\
\bottomrule()
\caption{\label{tbl-severity}Examples of outlier severities for
different days, where arrows indicate positive or negative
outliers}\tabularnewline
\end{longtable}

\hypertarget{visualising-detected-outliers-for-analysts}{%
\subsection{Visualising detected outliers for
analysts}\label{visualising-detected-outliers-for-analysts}}

There are multiple different visualisations that could be used to
present the information from Table~\ref{tbl-severity} to analysts.

To support on-the-day forecast adjustments, the simplest approach is as
a ranked \emph{alert list}, as exemplified by
Table~\ref{tbl-alert_list_ch3}. By presenting the alert list as a table
rather than a visualisation, this gives a clear, prioritised list of
tasks to complete. Here, the different colours show where the outlier
was detected: \textcolor{red}{red} showing the stations where the
outlier was detected, and \textcolor{cyan}{blue} showing other stations
in the same cluster likely to be affected. By displaying the severity
alongside the ranking, analysts are better able to prioritise their
adjustments. For example, an analyst may choose to only adjust the
forecasts for the top two outliers in Table~\ref{tbl-alert_list_ch3}, as
the third has a much lower severity in comparison.

\begin{table}

\end{table}

\hypertarget{tbl-alert_list_ch3}{}
\begin{longtable}[]{@{}
  >{\centering\arraybackslash}p{(\columnwidth - 6\tabcolsep) * \real{0.0787}}
  >{\centering\arraybackslash}p{(\columnwidth - 6\tabcolsep) * \real{0.0787}}
  >{\centering\arraybackslash}p{(\columnwidth - 6\tabcolsep) * \real{0.1654}}
  >{\raggedright\arraybackslash}p{(\columnwidth - 6\tabcolsep) * \real{0.6772}}@{}}
\toprule()
\begin{minipage}[b]{\linewidth}\centering
Rank
\end{minipage} & \begin{minipage}[b]{\linewidth}\centering
Severity
\end{minipage} & \begin{minipage}[b]{\linewidth}\centering
Direction of change
\end{minipage} & \begin{minipage}[b]{\linewidth}\raggedright
Cluster
\end{minipage} \\
\midrule()
\endfirsthead
\toprule()
\begin{minipage}[b]{\linewidth}\centering
Rank
\end{minipage} & \begin{minipage}[b]{\linewidth}\centering
Severity
\end{minipage} & \begin{minipage}[b]{\linewidth}\centering
Direction of change
\end{minipage} & \begin{minipage}[b]{\linewidth}\raggedright
Cluster
\end{minipage} \\
\midrule()
\endhead
1 & 0.892 & \(\uparrow\) &
\textcolor{red}{31104, 31115, 31129, 31217, 31219, 31222, $\hdots$} \\
2 & 0.828 & \(\uparrow\) & \textcolor{red}{32008, 32048} \\
3 & 0.347 & \(\uparrow\) &
\textcolor{red}{31000, }\textcolor{cyan}{31001, 31002, 31003, 31004, 31005, $\hdots$} \\
\(\vdots\) & \(\vdots\) & \(\vdots\) & \(\vdots\) \\
\bottomrule()
\caption{\label{tbl-alert_list_ch3}Example of ranked alert list for
30/03/2018}\tabularnewline
\end{longtable}

\begin{figure}

{\centering \includegraphics{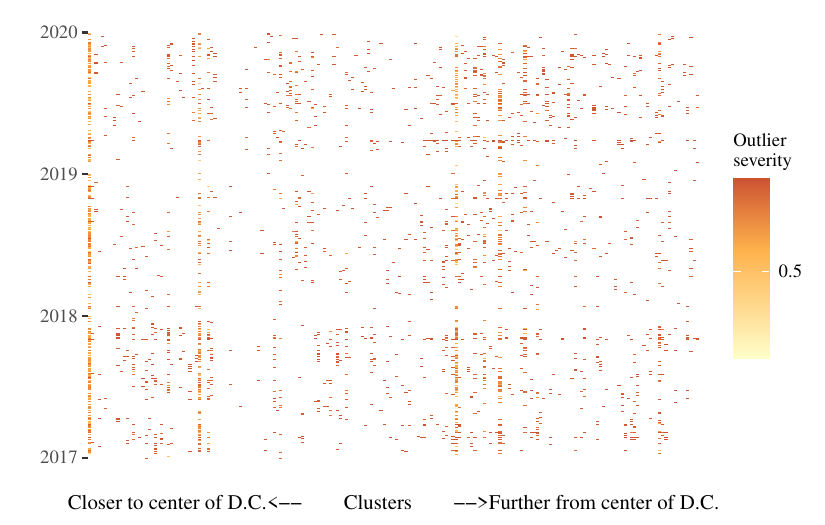}

}

\caption{\label{fig-cluster_heatmap}Outlier severity for each cluster
between 2017 and 2019. Figure originally included in
\citet{Rennie_thesis}.}

\end{figure}

To give a wider view of how outliers have occurred and to account for
them in the historic data, the severity per cluster can be visualised in
a spatiotemporal heatmap as exemplified in
Figure~\ref{fig-cluster_heatmap}. This figure shows the severity of
detected outliers over time for every cluster, where clusters are
arranged from left to right by nearest to furthest from the centre of
Washington D.C. The order of the clusters along the x-axis could also be
arranged to highlight further spatial patterns e.g.~from North to South.
This type of visualisation can help to identify large-scale patterns in
the outliers. For example, knowing for which times of year or days of
the week outliers are more likely to be detected can help to steer the
attention of analysts. Similarly, dedicated analysts may be assigned to
monitor dedicated clusters, and this helps to identify which clusters
may need more manual adjustments.

In addition to identifying patterns in Figure~\ref{fig-cluster_heatmap},
analysts could utilise this figure to find changes in the outlier
patterns. For example, changes in the distribution of outliers along the
x-axis would suggest a change in the spatial demand pattern, and could
indicate a refresh of the clustering process is required. Changes in the
distribution of outliers along the y-axis may indicate more or fewer
outliers over time or, if the new pattern is regular, suggest that a
variable is missing from the demand forecast. Changes in the colour of
points may indicate that outliers are becoming more (or less) severe,
i.e.~demand is becoming less (or more) predictable, which could prompt a
review of the forecasting and optimisation approached. For example, it
could indicate that a more robust approach to optimisation is required
if outliers are occurring more frequently or with higher severity.

\hypertarget{sec-discussion}{%
\section{Discussion}\label{sec-discussion}}

In this section, we discuss patterns in the outliers detected in the
Capital Bikeshare data, and suggest potential explanations for their
causes.

\begin{figure}

{\centering \includegraphics{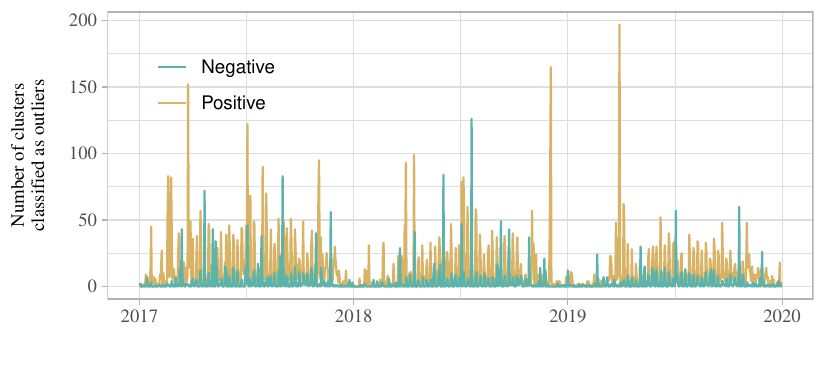}

}

\caption{\label{fig-pos_neg_line}Positive and negative outliers. Figure
originally included in \citet{Rennie_thesis}.}

\end{figure}

Figure~\ref{fig-pos_neg_line} visualises the number of positive and
negative outliers over the days of the time span recorded in the data
set. By visualising the positive and negative outliers jointly, analysts
can immediately see that (i) negative outliers typically affect far
fewer clusters than positive outliers; (ii) the spikes where outlier
demand affects a large number of clusters do not occur at the same time
for positive and negative outliers; and (iii) the seasonal patterns in
the detected outliers are not the same for positive and negative
outliers. This can aid analysts in their predictions of outliers: though
positive outliers are always more common, the proportion of outliers
that are negative changes throughout the year. Though the largest
negative outliers occur in summer, the transition between summer and
winter seems particularly affected by negative outliers, with May and
October exhibiting the highest proportion of negative outliers at around
33\%.

The increased occurrence of positive outliers can be explained by the
fact that demand is bounded below by zero, and in many cases the mean
usage pattern is close to zero, such that negative demand is
unobservable. This may motivate a transformation of the data before
outlier detection is performed -- see Appendix \ref{sec-app-skew} for
further details.

Outliers occur independently in different clusters. In fact, only four
days observe outliers in more than 125 of the 195 clusters -- one
negative and three positive. The three positive outliers occur on 25
March 2017, 3 December 2018, and 30 March 2019. Explanations from events
arise for two of these dates: 3 December 2018 relates to the funeral of
George H. W. Bush, and a NATO protest occurred in Washington D.C. on 30
March 2019. 25 March 2017 and 30 March 2019 were both warm days, and the
last Saturday in March - perhaps suggesting that the definition of
\emph{summer} should be from the last Saturday in March, rather than
April 1.

It is interesting to note that the next most widespread positive outlier
relates to Independence Day in 2017. Independence Day was detected as a
positive outlier in 123, 80, and 35 clusters in 2017, 2018, and 2019.
However it was detected as a negative outlier in 46, 47, and 57 stations
respectively. The date of the widespread negative outlier is 21 July
2018 which relates to one of the worst storms Washington D.C. has seen.
Further discussion of how weather is related to outliers can be found in
Section \ref{sec-weather}.

\hypertarget{spatiotemporal-patterns-in-detected-outliers}{%
\subsection{Spatiotemporal patterns in detected
outliers}\label{spatiotemporal-patterns-in-detected-outliers}}

In this section, we analyse the detected outliers and consider spatial
and temporal patterns within the outliers.

\hypertarget{temporal-patterns.}{%
\paragraph{Temporal patterns.}\label{temporal-patterns.}}

Even after accounting for the lower means and reduced inter-daily
variance of the winter months, we detect fewer outliers in winter
(indicated by the two horizontal white bars in
Figure~\ref{fig-cluster_heatmap}). Otherwise, we find no clear
systematic temporal patterns to the detected outliers. Appendix
\ref{sec-app-preproc_disc} provides additional discussion on the
temporal aspects of the detected outliers, including the visible
temporal patterns in the outliers when we fail to account for temporal
patterns in the forecasting step.

\begin{figure}

{\centering \includegraphics{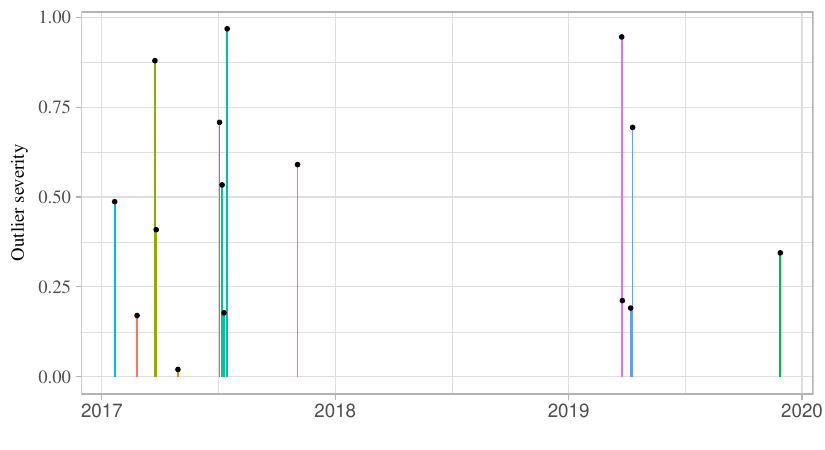}

}

\caption{\label{fig-outliers_temporal_clusters}Exemplified severity for
outliers detected in one cluster, showing temporal clustering of
outliers with each colour representing one of nine clusters. Figure
adapted from \citet{Rennie_thesis}.}

\end{figure}

However, Figure~\ref{fig-outliers_temporal_clusters} shows that although
outliers can sometimes occur as one-off events, they are also quite
likely to occur in temporal clusters. Therefore, once an outlier has
been identified, the information can be used to support adjustments to
planning in the subsequent days.

\hypertarget{spatial-patterns.}{%
\paragraph{Spatial patterns.}\label{spatial-patterns.}}

Next, we discuss spatial patterns in the detected outliers and consider
the relationship between outliers in pick-up and drop-off usage
patterns. Figure~\ref{fig-cluster_heatmap} shows that the cluster which
is formed around the centre of Washington D.C. (indicated by the first
column on the left) experiences more frequent and higher probability
outliers. Otherwise, there is little geographic pattern to how often
outliers occur in terms of distance from the centre.

\begin{figure}

{\centering \includegraphics{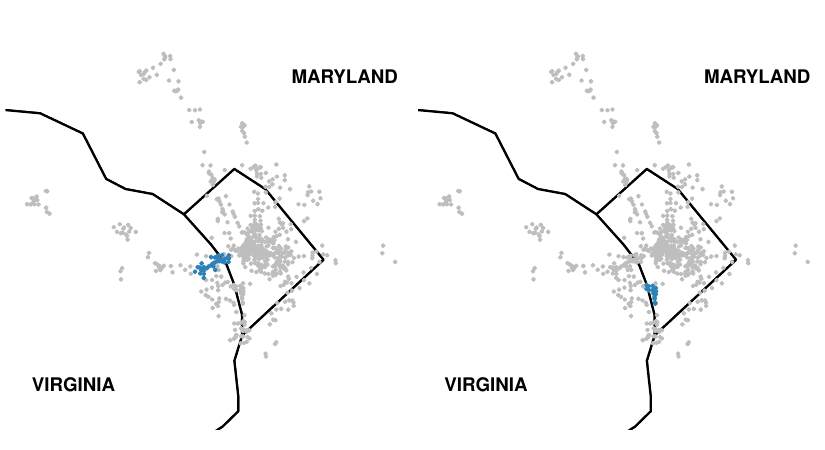}

}

\caption{\label{fig-cluster_hm_highlight}Two (non-central D.C.) clusters
which exhibit higher numbers of outliers. Figure originally included in
\citet{Rennie_thesis}.}

\end{figure}

Two other clusters besides the central D.C. cluster exhibit a higher
number of outliers with higher severity than other clusters.
Figure~\ref{fig-cluster_heatmap} indicates these by darker vertical
lines. These clusters are highlighted in
Figure~\ref{fig-cluster_hm_highlight}. These clusters are both fairly
close to the centre of Washington D.C., and are close by the two main
bridges across the Potomac River into the centre. The stations in these
clusters are also situated close to The Pentagon, Arlington National
Cemetery, and Ronald Reagan Washington National Airport. Therefore these
clusters are likely to experience business commuter demand, tourist
demand, and potentially also airport travellers i.e.~have multiple
sources of outlier demand.

\begin{figure}

{\centering \includegraphics{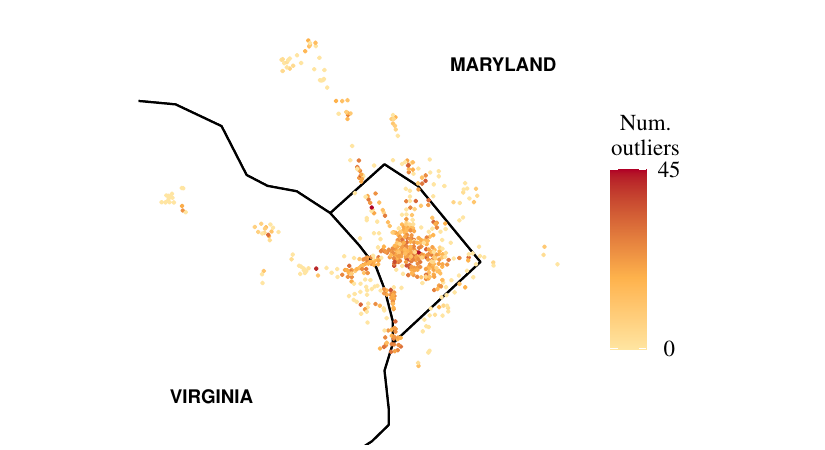}

}

\caption{\label{fig-cluster_num_outs}Number of days each station was
classified as an outlier between 2017-2019. Figure originally included
in \citet{Rennie_thesis}.}

\end{figure}

We also consider the frequency of outliers on the station level.
Figure~\ref{fig-cluster_num_outs} shows the number of days that each
individual station was classified as an outlier between 2017 and 2019.
Stations where no outliers were detected are not shown. Outliers are
more commonly detected in stations nearer the centre of D.C.

We analyse the differences in the spatial patterns of the outliers
detected in pick-up and drop-off usage patterns. For this, we use the
clustering based on the overall usage patterns as it allows direct
comparison of outliers in different clusters. Subsequently, we apply the
outlier detection procedure separately to the pick-up and drop-off usage
patterns. This enables us to isolate how the detected outliers and their
severities change when the stations in each cluster remain constant.
Additional discussion of the spatiotemporal patterns of the outliers can
be found in Appendix \ref{sec-app-add_disc}, including further
visualisations that analysts may use to identify such patterns to aid in
decision-making.

\begin{figure}

{\centering \includegraphics{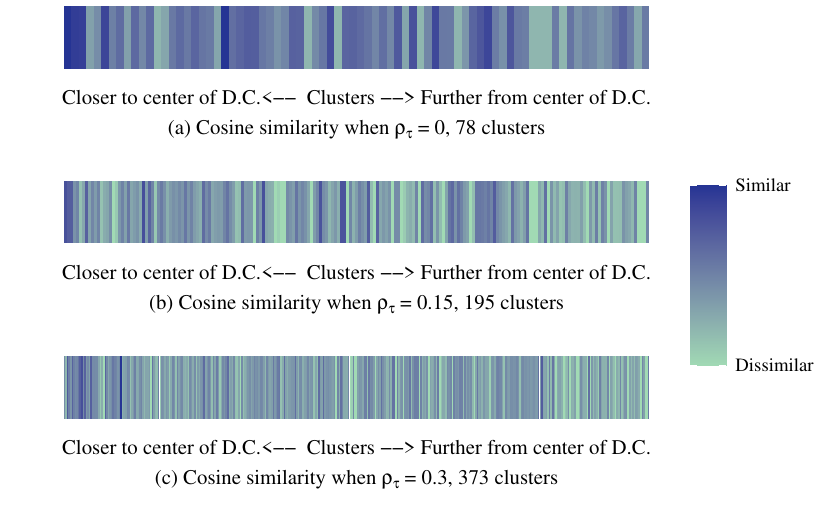}

}

\caption{\label{fig-cosine_start_end}Cosine similarity between outliers
detected in pick-up and drop-off usage patterns under different
correlation thresholds. Figure originally included in
\citet{Rennie_thesis}.}

\end{figure}

To formally compare the output of the outlier detection procedure for
pick-up and drop-off usage patterns, we use cosine similarity
\citep{Leydesdorff2005}. That is, the cosine of the angle between two
vectors, where 0 represents complete dissimilarity, and 1 complete
similarity. Figure~\ref{fig-cosine_start_end}(a) provides the cosine
similarity between clusters i.e.~the cosine similarity of the vector of
outlier severities for those detected in pick-up stations over the three
year period, and that for drop-off stations.

Figure~\ref{fig-cosine_start_end}(a) shows that outliers detected in
pick-up and drop-off stations are fairly similar, although this changes
depending on the correlation threshold used in the clustering step. As
the correlation threshold ranges from 0 to 0.3, the average cosine
similarity ranges from 0.69 to 0.44. As the correlation threshold
increases, the number of clusters increases. Therefore, when we look at
outliers on a small cluster or station level, there is less similarity
between pick-ups and drop-offs. However, when the clusters are larger
and the outliers aggregated, there is a clearer pattern between pick-ups
and drop-offs. Further, this similarity is not uniform across the
different clusters - those closer to the centre of D.C. have a higher
cosine similarity. That is, the closer a cluster is to the centre of
D.C., the more likely it is that if a day is a pick-up station cluster
outlier, it will also be a corresponding drop-off station cluster
outlier. We did not find any temporal patterns in the comparison of
pick-up and drop-off outliers.

\hypertarget{sec-weather}{%
\subsection{Weather as an explanatory factor for demand
outliers}\label{sec-weather}}

It is widely acknowledged that weather can be used as a predictor for
average bike-sharing demand \citep{Lin2020}. Therefore, we examine
whether extreme temperature or rainfall drive extreme changes in demand
i.e.~outliers. To that end, we analyse weather data obtained from
\citet{VisCros} and investigate whether weather can be used to explain
and eventually predict the outliers in demand. The data is included in
Appendix \ref{sec-app-weather}.

\begin{figure}

{\centering \includegraphics{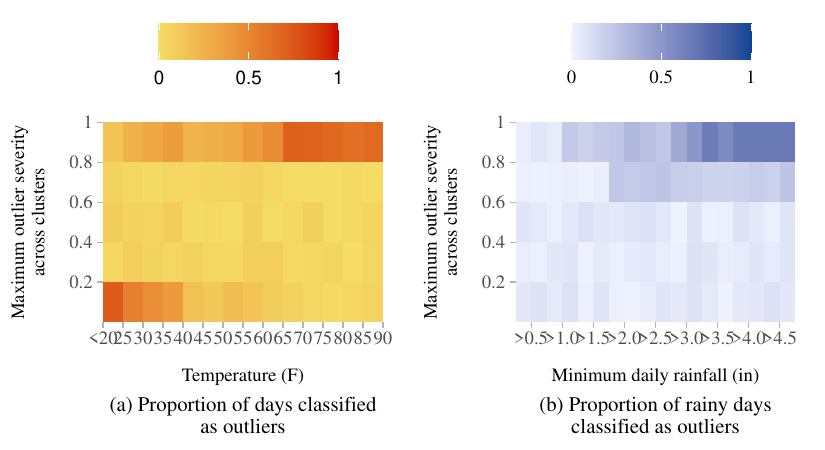}

}

\caption{\label{fig-weather_outliers}Severity of outliers at different
temperatures and precipitation levels. Figure originally included in
\citet{Rennie_thesis}.}

\end{figure}

Figure~\ref{fig-weather_outliers}(a) shows the proportion of days in
each temperature range that have a maximum outlier severity within each
severity range. Higher temperatures (between 70 and 90 \(\circ\) F)
result in higher severity outliers, indicated by the red region in the
top right of Figure~\ref{fig-weather_outliers}(a). The red region in the
bottom left shows that a high proportion of days with a very low average
daily temperatures, \(\leq\) 25 \(\circ\) F, are classified as outliers.
However, these outliers typically have a low severity. This can be
explained by these outliers being negative demand outliers -- and low
temperatures typically occur in winter, when demand is already low.

In addition to temperature, we also expect precipitation levels to
affect demand for bike-sharing. As we expect increased rainfall to have
a negative impact of usage, we consider only the severity of negative
demand outliers here. Figure~\ref{fig-weather_outliers}(b) shows the
proportion of days with a minimal level of precipitation which were
classified as outliers with some severity. Higher rainfall generally
results in higher likelihood of the day being classed as a negative
demand outlier. When precipitation levels are especially high, the
outliers that are detected also tend to have higher severities.

There are likely many other factors which cause and influence outlier
demand in the Capital Bikeshare network. For example, \citet{Ma2015}
have previously linked usage of bike-sharing stations to usage of
Metrorail services. We anticipate that cancellations, or short-term
changes to Metrorail services may also generate outlier demand for
bike-sharing. However, due to lack of available of data on such
cancellations, we leave this to future research.

\hypertarget{sec-ch3_conclusion}{%
\section{Conclusion}\label{sec-ch3_conclusion}}

In this paper, we identified temporal patterns in the Capital Bikeshare
data set and applied a combination of functional regression and temporal
partitioning to remove such trends and obtain a homogeneous data set. We
also accounted for spatial patterns in temporal usage by clustering
together similar stations. By basing our clustering algorithm on a
combination of geographical knowledge, and similarity of usage patterns,
we were able to identify how usage changes as stations get further away
from the city centre. Throughout this paper, we have presented
visualisations to illustrate our findings and provided detailed
descriptions of how such visualisations may be used by analysts to aid
in their decision making.

Our in-depth study of detected outliers showed that not all stations are
equally prone to outliers - those closest to the centre of D.C. exhibit
far more outlying demand. This is also true for known outlier days
e.g.~national holidays such as July 4, where some clusters of stations
exhibit increased demand and others decreased. For forecasting and
planning purposes, this knowledge is highly important since outlier
demand changes not only the magnitude of demand, but also the spatial
distribution of where customers go. In terms of rebalancing bikes at
stations, this could have a large impact on the efficiency of the
schedule. Further, we also showed that outliers are more likely to occur
in the summer months (even after accounting for increased usage and
usage variability), suggesting rebalancing needs to be more reactive in
the summer months.

Our analysis of weather patterns showed that outliers are more likely to
occur when the weather conditions are more extreme. Both temperature and
precipitation were found to have an impact on demand - with excessively
high precipitation or very low temperatures causing negative demand
outliers, and high temperatures causing positive demand outliers.

Further research is needed to evaluate the effects that identifying and
correcting for outliers may have on revenue and planning in the
bike-sharing domain. This could include considering the impact of
forecast accuracy on re-allocation and revenue, and the cost-benefit
relationship that may result from making a change to the forecast to
account for outlier demand. The method outlined in this paper could be
used to generate an outlier \emph{alert}, to notify Capital Bikeshare
when the rebalancing policy for a given day is non-optimal. Online
detection of outlier demand would allow re-allocation of bikes during
the day and, as such, further analysis of how these alerts could be
deployed in an automated system, and how they may affect the complexity
of the routing problem, is needed.

A further extension could consider how outlier detection may be applied
to dockless bike-sharing systems - where users may pick-up or drop-off a
bike anywhere rather than at dedicated stations. Similar methodology
could be applied in this case, though an additional pre-processing step
would be required - where regions are defined. All pick-ups or drop-offs
within a single region would then be treated in the same way as a single
station. These regions may be defined based on knowledge of the
underlying city geography, or by applying spatial clustering methods.

\hypertarget{data-availability-statement}{%
\section{Data availability
statement}\label{data-availability-statement}}

The datasets analysed during this study are available from Capital
Bikeshare in a public repository that does not issue datasets with DOIs
at
\href{https://s3.amazonaws.com/capitalbikeshare-data/index.html}{s3.amazonaws.com/capitalbikeshare-data/index.html}.
Further information on any pre-processing performed by Capital Bikeshare
is available at
\href{https://ride.capitalbikeshare.com/system-data}{ride.capitalbikeshare.com/system-data},
and the license agreement at
\href{https://ride.capitalbikeshare.com/data-license-agreement}{ride.capitalbikeshare.com/data-license-agreement}.

The paper at hand was included as Chapter 4 in the doctoral thesis
\citet{Rennie_thesis}. The figures within this paper are also included
in, or adapted from, \citet{Rennie_thesis}.

\newpage{}

\appendix

\hypertarget{additional-analysis}{%
\section{Additional analysis}\label{additional-analysis}}

\hypertarget{forecasting-baseline-demand}{%
\subsection{Forecasting baseline
demand}\label{forecasting-baseline-demand}}

\hypertarget{sec-app-preproc}{%
\subsubsection{Temporal partitioning}\label{sec-app-preproc}}

\begin{figure}

{\centering \includegraphics{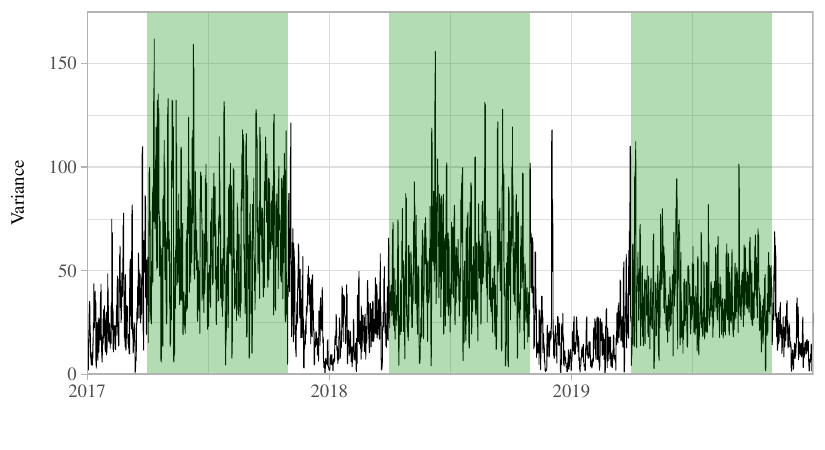}

}

\caption{\label{fig-variance_rentals}Variance of usage patterns with
summer months highlighted in green. Figure originally included in
\citet{Rennie_thesis}.}

\end{figure}

In Section \ref{sec-st_patterns}, for the purposes of temporal
partitioning of data, we define summer to be the months April through
October. Winter is therefore November through March. The partitioning is
chosen to give constant variance within a partition whilst also ensuring
there is a sufficient number of observations within each partition to
make outlier detection feasible. Figure~\ref{fig-variance_rentals} shows
the rolling daily variance for station 31203, with the summer months
highlighted.

\begin{figure}

{\centering \includegraphics{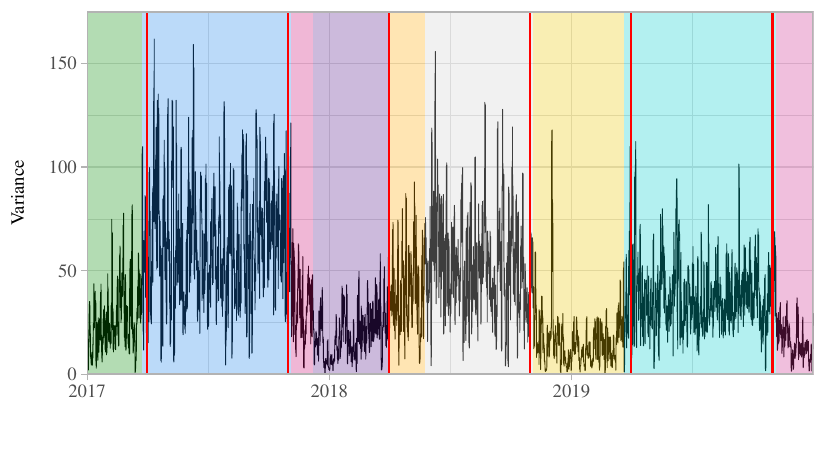}

}

\caption{\label{fig-cpt}Changepoints in variance of usage patterns.
Figure originally included in \citet{Rennie_thesis}.}

\end{figure}

The variance of winter is not constant, being slightly higher in the
months that border the summer season. Further partitioning could be
carried out e.g.~partition by month. However, this results in much less
data within each partition, which then makes outlier detection more
difficult. When applying binary segmentation changepoint detection
\citep{Scott1974} to identify the partitions with different levels of
variance, the algorithm returns 8 changepoints: 24 March 2017, 4 Nov
2017, 6 Dec 2017, 31 Mar 2018, 24 May 2018, 4 Nov 2018, 20 Mar 2019, and
4 Nov 2019. These are highlighted in Figure~\ref{fig-cpt}. These are
relatively close to our pre-defined summer and winter partitions
(indicated by red vertical lines in Figure~\ref{fig-cpt}.

If there are already pre-defined seasons in use for planning purposes,
these may be more appropriate.

\hypertarget{sec-app-model_comp}{%
\subsubsection{Functional regression model
comparison}\label{sec-app-model_comp}}

In this section, we perform model comparison for the functional
regression model used to account for different daily trends as detailed
in Section \ref{sec-st_patterns}, Equation~\ref{eq-funcreg}.

We use the \textbf{Cross-Validated Mean Integrated Squared Error}
(CV-MSE) to determine the best-fitting model. The CV-MSE is given by:

\begin{equation}
    CV\mbox{-}MSE = \frac{1}{N}\sum_{n=1}^{N} \int (x_{n,s}(t) - \hat{x}_{n,s}(t) ) dt, 
\end{equation}

where \(\hat{x}_{n,s}(t)\) is the prediction for the \(n^{th}\) daily
rental pattern at the station \(s\), under the model fitted to all but
the \(n^{th}\) rental pattern. The model which produces the lowest
CV-MSE is chosen as the best fitting. Unlike other model selection
criterion such AIC, CV-MSE does not take into account the number of
parameters. The CV-MSE for each of the 8 models considered is shown in
the table below, for the stations in the cluster discussed in Section
\ref{sec-outliers_method}.

\begin{table}[!ht]
\centering
\resizebox{1\textwidth}{!}{\begin{tabular}{c|ccc|ccccccccc}
\hline \hline
\multirow{2}{*}{Model} & \multicolumn{3}{c}{Factors}          & \multicolumn{9}{c}{Station Number}                                                                                                                           \\ \cline{2-13}
                       & Day        & Month      & Year       & 31303          & 31308          & 31309          & 31315          & 31316          & 31317          & 31319          & 32014          & 32040          \\ \hline
1                      &            &            &            & 135.90         & 72.98          & 16.04          & 25.80          & 15.29          & 27.96          & 34.27          & 52.42          & 22.75          \\
2                      & \cmark &            &            & 120.38         & 60.17          & 16.04          & 25.64          & 14.29          & 27.71          & 33.43          & 50.80          & 22.78          \\
3                      &            & \cmark &            & 98.01          & 59.65          & 12.67          & 18.82          & 11.93          & 19.57          & 25.78          & 37.97          & 17.09          \\
4                      &            &            & \cmark & 133.96         & 71.89          & 15.49          & 25.70          & 14.86          & 28.03          & 33.97          & 48.01          & 22.63          \\
5                      & \cmark & \cmark &            & 82.40          & 46.50          & 12.61          & 18.65          & 10.83          & \textbf{19.20} & 24.80          & 36.07          & 17.01          \\
6                      &            & \cmark & \cmark & 96.59          & 58.52          & 12.07          & 18.79          & 11.52          & 19.61          & 25.44          & 33.34          & 16.95          \\
7                      & \cmark &            & \cmark & 119.04         & 59.02          & 15.47          & 25.63          & 13.87          & 27.77          & 33.13          & 46.28          & 22.66          \\
8                      & \cmark & \cmark & \cmark & \textbf{80.85} & \textbf{45.29} & \textbf{12.00} & \textbf{18.62} & \textbf{10.39} & 19.23          & \textbf{24.46} & \textbf{31.32} & \textbf{16.87} \\ \hline \hline
\end{tabular}}
\caption{Cross validated mean square error for functional regression model comparison applied to unpartitioned data}
\label{tbl-model_comp}
\end{table}

In most cases, the model which achieves the minimum mean squared error
is model 8, which includes all three factors (day, week, and year).
However, model 5 (day and month) also produces very similar results.

\hypertarget{sec-app-ridges}{%
\subsubsection{Distribution of residuals}\label{sec-app-ridges}}

\begin{figure}

{\centering \includegraphics{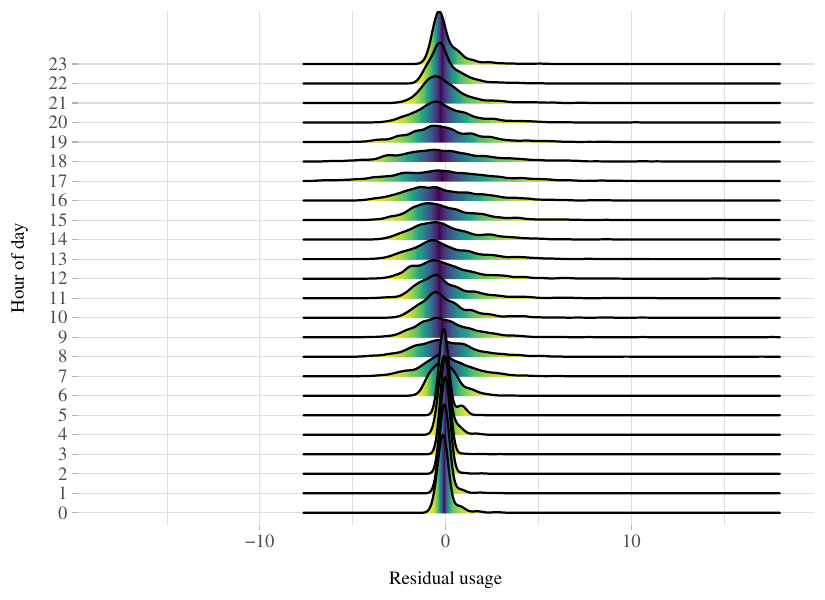}

}

\caption{\label{fig-residuals_ridges}Distribution of residual usage for
each hour of the day for station 31005. Figure originally included in
\citet{Rennie_thesis}.}

\end{figure}

Figure~\ref{fig-residuals_ridges} shows the distribution of the
residuals for each hour of the day for station 31005 -- see also
Figure~\ref{fig-residuals}. The core of the distribution is symmetric
around zero, but the tails of the distribution are positively skewed.

\hypertarget{sec-app-skew}{%
\subsubsection{Accounting for skewness}\label{sec-app-skew}}

\begin{figure}

{\centering \includegraphics{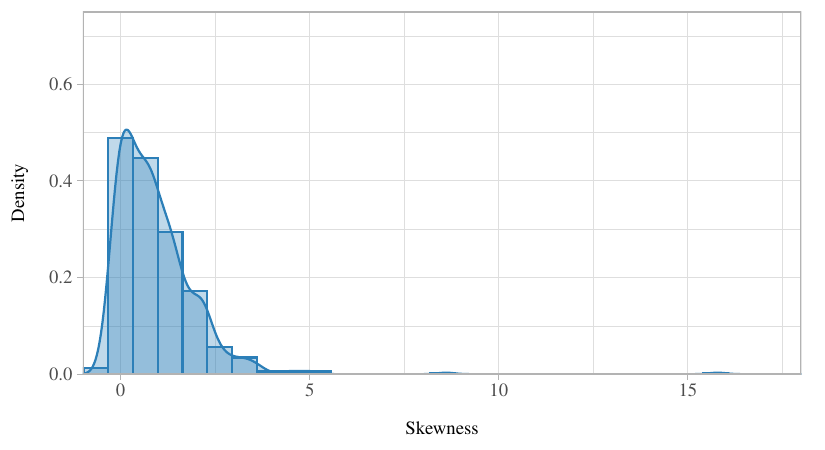}

}

\caption{\label{fig-skew_dist}Distribution of skewness of distributions
of total daily usage across all stations. Figure originally included in
\citet{Rennie_thesis}.}

\end{figure}

\begin{figure}

{\centering \includegraphics{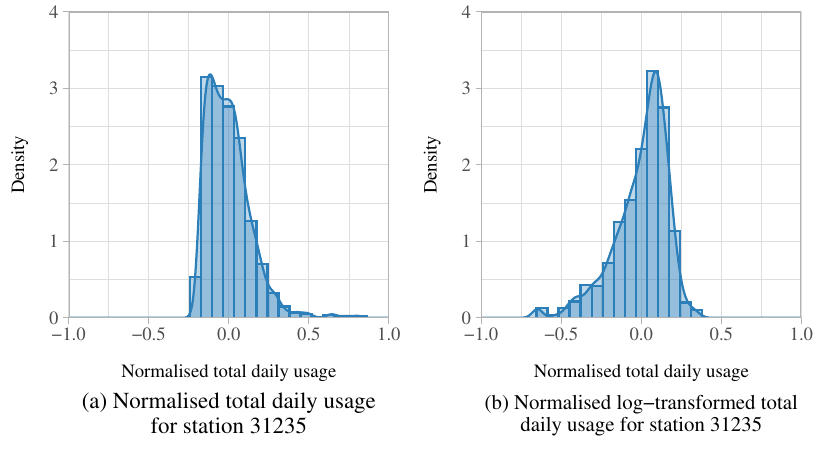}

}

\caption{\label{fig-usage_hist}Distribution of total daily usage for
station 31235. Figure originally included in \citet{Rennie_thesis}.}

\end{figure}

Figure~\ref{fig-usage_hist}(a) shows the distribution of the normalised
total daily usage for station 31235, which exhibits positive skew. Not
all stations exhibit such positively skewed distributions -- see
Figure~\ref{fig-skew_dist}.

The distributions of total daily usage have a skewness lying between
-0.4 and 15.8, with the median skewness across all stations being 0.71.
Larger positive skew is more common in stations where mean usage is very
low, and since demand is bounded below by zero, only increases in demand
are observed. This results in more \emph{positive} outliers than
\emph{negative} (see Section \ref{sec-discussion}). Given that most
stations exhibit slight positive skew, it may be desirable to transform
the data before performing outlier detection. To account for the skew,
the rental patterns can first be transformed e.g.~with a logarithmic
transform. However, this is not applicable to all stations (as some are
already negatively skewed) and can result in a negatively skewed
distribution -- Figure~\ref{fig-usage_hist}.

When applying the outlier detection procedure to the untransformed data,
the fraction of positive outliers is consistently higher than the
fraction of negative outliers. On average, 78\% of outliers are
positive. That is, outliers are more likely to be caused by increased
demand than decreased. This is easily explained by the fact that demand
is bounded below by zero, and in many cases the mean usage pattern is
close to zero, such that negative demand is unobservable. Applying a
logarithmic transformation before carrying out the outlier detection
results in around 60\% of outliers being positive.

\begin{figure}

{\centering \includegraphics{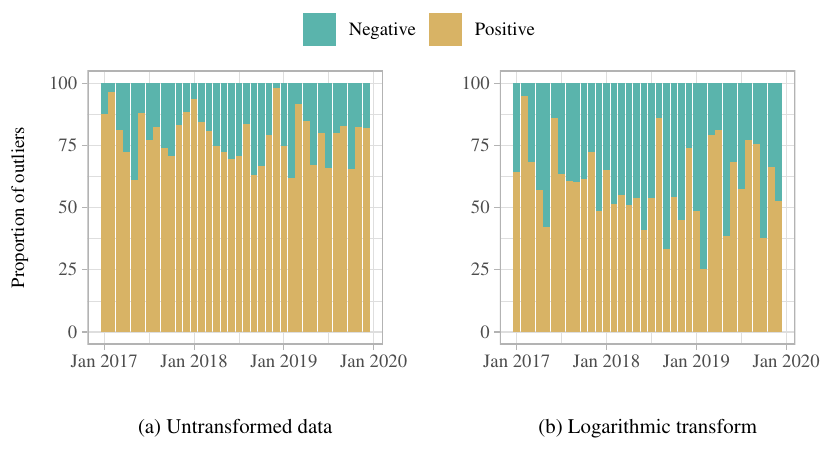}

}

\caption{\label{fig-pos_neg_transform}Fraction of outliers that are
positive and negative, before and after applying a logarithmic
transformation. Figure originally included in \citet{Rennie_thesis}.}

\end{figure}

\hypertarget{sec-app-acf}{%
\subsubsection{Inter-daily autocorrelation}\label{sec-app-acf}}

\begin{figure}

{\centering \includegraphics{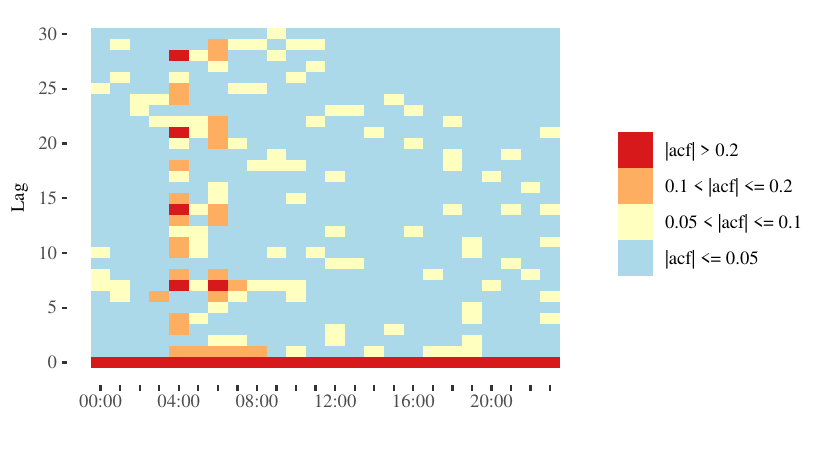}

}

\caption{\label{fig-acf_heatmap}Inter-daily autocorrelations of
residuals for station 31005. Figure originally included in
\citet{Rennie_thesis}.}

\end{figure}

Figure~\ref{fig-acf_heatmap} shows the inter-daily autocorrelations
between the residual patterns for different days, at each hour. The
early hours of the morning - especially at 04:00 and 06:00 - exhibit
some autocorrelation of lag 7 i.e.~weekly. A functional ARIMA model
could be fitted to remove the autocorrelation. However, as it only
affects a so few hours of the day, we do not investigate this further
here.

\hypertarget{using-spatial-patterns-to-cluster-stations}{%
\subsection{Using spatial patterns to cluster
stations}\label{using-spatial-patterns-to-cluster-stations}}

\hypertarget{sec-app-cluster_params}{%
\subsubsection{Effect of parameter choices on
clustering}\label{sec-app-cluster_params}}

\begin{figure}

{\centering \includegraphics{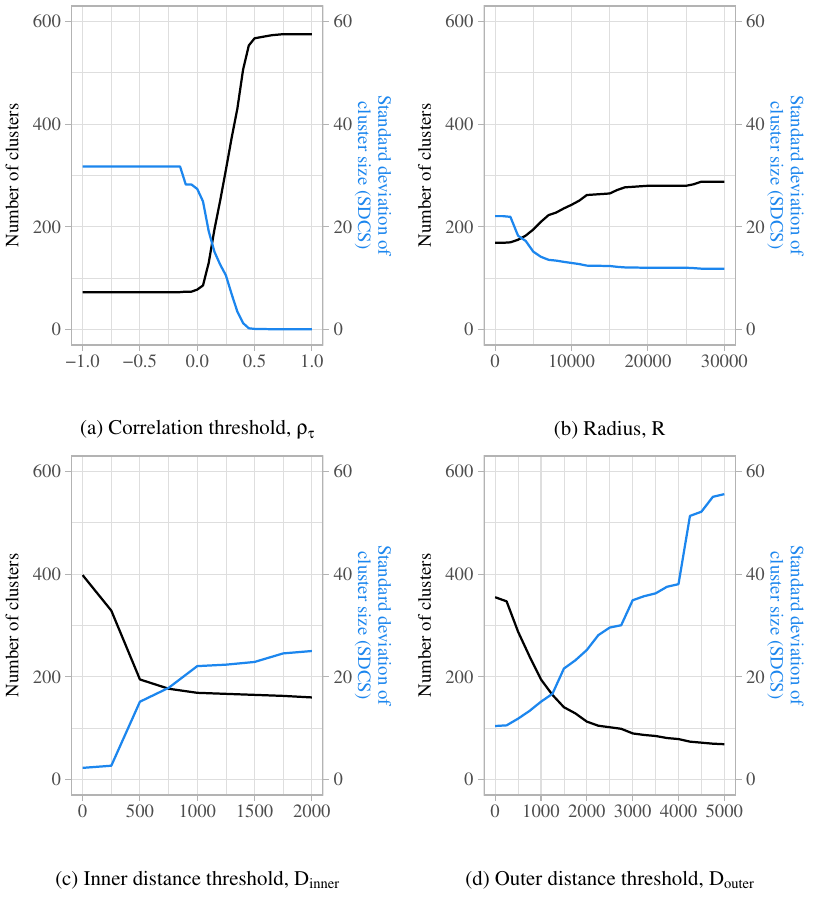}

}

\caption{\label{fig-mst_params}Cluster sensitivity to parameter changes
when other parameters remain fixed at \(\rho_{\tau}\)=0.15, \(R\) =
5000m, \(D_{inner}\) = 500m, and \(D_{outer}\) = 1000m. Figure
originally included in \citet{Rennie_thesis}.}

\end{figure}

Our clustering method is tuned using four parameters: the correlation
threshold \(\rho\), as well as distance metrics introduced in Section
\ref{sec-clustering_ch3}. We now evaluate the sensitivity of changing
these parameters on (i) the number of clusters obtained, and (ii) the
standard deviation in cluster sizes (SDCS) \citep{Lin2019}. The SDCS is
given by:

\begin{equation}
    SDCS = \sqrt{\frac{1}{K-1} \sum_{k=1}^K \left(S_k - \frac{S}{K}\right)^2},
\end{equation}

where \(S\) is the number of stations, \(K\) is the number of clusters,
\(S_k\) is the number of stations in cluster \(k\). The SDCS quantifies
a measure of the balance of the different cluster sizes. We do not seek
to minimise nor maximise the SDCS -- since choosing extreme parameter
values trivially creates clusters of size 1 or one giant cluster.

Figure~\ref{fig-mst_params} shows the change in number of clusters and
SDCS as we vary parameter values. There is an inverse relationship
between the number of clusters and the SDCS across all four variables.
While an increase in either correlation threshold or radius results in a
decrease of SDCS, increasing either of the distance thresholds increases
the SDCS. In order to achieve a balance between number of clusters and
SDCS, we choose parameter values close to the intersection* of the two
lines. This results in a correlation threshold of between 0 and 0.4; a
radius between 5,000m and 10,000m; an inner distance threshold between
500m and 1,000m, and an outer distance threshold of approximately
1,000m.

\hypertarget{sec-app-nmi_def_ch3}{%
\subsubsection{Normalised Mutual
Information}\label{sec-app-nmi_def_ch3}}

For a graph containing \(M\) stations, the mutual information between
two clusterings \(\mathcal{A}\) and \(\mathcal{B}\) of the \(M\) nodes
in the graph is defined as:

\begin{equation}
    I(\mathcal{A},\mathcal{B}) = \sum_{a=1}^{|\mathcal{A}|} \sum_{b=1} ^{|\mathcal{B}|} \frac{|\mathcal{A} \cap \mathcal{B}|}{M} \mbox{log} \left(|\mathcal{A} \cap \mathcal{B}|\frac{M}{M_a M_b}\right),
\end{equation} where \(M_a\) is the number of nodes in the \(a^{th}\)
cluster of clustering \(\mathcal{A}\), and similarly for \(M_b\). The
\textbf{normalised mutual information (NMI)} between two clusterings is
defined as \citep{Amelio2015}: \begin{equation}
    NMI(\mathcal{A},\mathcal{B}) = \frac{2I(\mathcal{A},\mathcal{B})}{H(\mathcal{A})+H(\mathcal{B})},
\end{equation} where \(H(\mathcal{A})\) is the entropy (a measure of
uncertainty) defined as: \begin{equation}
    H(\mathcal{A}) = - \sum_{a=1}^{|\mathcal{A}|} \frac{M_a}{M} \mbox{log} \left(\frac{M_a}{M}\right).
\end{equation}

\(NMI(\mathcal{A},\mathcal{B})\) = 1 if \(\mathcal{A}\) and
\(\mathcal{B}\) are identical, and 0 if they are completely different.

\hypertarget{sec-app-add_disc}{%
\subsection{Additional Discussion}\label{sec-app-add_disc}}

\hypertarget{sec-app-preproc_disc}{%
\subsubsection{Effects of data temporal patterns on outlier
detection}\label{sec-app-preproc_disc}}

\begin{figure}

{\centering \includegraphics{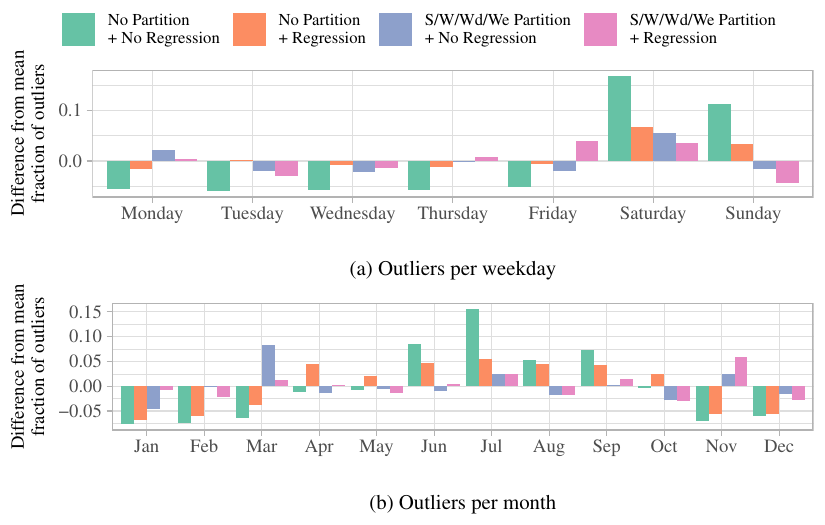}

}

\caption{\label{fig-regression_output}Fraction of outliers occurring on
each day of the week and month of the year, with and without applying
functional regression model. Figure originally included in
\citet{Rennie_thesis}.}

\end{figure}

In Section \ref{sec-st_patterns}, we outlined two steps (functional
regression and temporal partitioning) that could be undertaken to
account for different patterns in the data. Here, we consider how the
inclusion of these steps affects the outcome of the outlier detection.
For a homogeneous data set, we would expect approximately equal numbers
of outliers detected on each day of the week, and month of the year.
Figure~\ref{fig-regression_output} shows the difference between the mean
fraction of outliers per day (or month) and fraction of outliers which
are observed on each day of the week (or month).

The results are shown for the case where the (i) there is no accounting
for temporal patterns; (ii) the regression model is applied with no
partitioning; (iii) only partitioning is applied with no regression, and
(iv) both regression and partitioning is applied. When we do not account
for temporal patterns in the data, we detect far more outliers on
weekends and in the summer months. Including the regression step
(without partitioning) improves this imbalance somewhat. When the data
has been partitioned, regression makes little difference to the
proportion of outliers detected on each day or month. Although we
partition the data to account for different variance, this implicitly
takes care of differences in mean between the same groups. As there is
little difference in mean trend between days or months in the same
groups. partitioning with or without regression gives similar results.

\hypertarget{sec-app-weather}{%
\subsubsection{Weather as an explanatory factor for demand
outliers}\label{sec-app-weather}}

\begin{figure}

{\centering \includegraphics{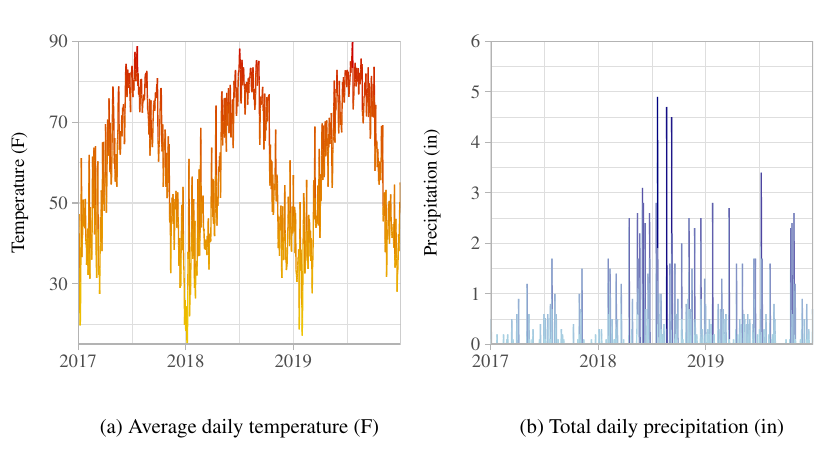}

}

\caption{\label{fig-app_weather}Weather data obtained from Visual
Crossing for 2017 - 2019. Figure originally included in
\citet{Rennie_thesis}.}

\end{figure}

Figure~\ref{fig-app_weather} shows the weather data used for analysis in
Section \ref{sec-weather}.

\clearpage

  \bibliography{references.bib}

\end{document}